\documentclass[10pt,journal,compsoc]{IEEEtran}
\usepackage{amsmath}
\usepackage{graphicx}
\usepackage{subfigure}
\usepackage{stfloats}
\usepackage{booktabs}
\usepackage{algorithm}
\usepackage{algpseudocode}
\usepackage{multirow}
\usepackage[switch]{lineno}
\usepackage{lineno}
\usepackage{amssymb}
\usepackage{booktabs}
\usepackage{multirow}
\usepackage{makecell}
\usepackage{tabularx}
\usepackage[font=normal]{caption}
\usepackage{float}

\usepackage[capitalize]{cleveref}
\crefname{section}{Sec.}{Secs.}
\Crefname{section}{Section}{Sections}
\Crefname{table}{Table}{Tables}
\crefname{table}{Tab.}{Tabs.}
\newcommand{\etal}{{\emph{et al.}}}

\usepackage[dvipsnames]{xcolor}
\definecolor{rc}{rgb}{0,0,0}
\ifCLASSOPTIONcompsoc
  \usepackage[nocompress]{cite}
\else
  \usepackage{cite}
\fi

\ifCLASSINFOpdf
\else

\fi

\begin{document}

\title{PNR: Physics-informed Neural Representation for
high-resolution LFM reconstruction}

\author{ Jiayin~Zhao, Zhifeng~Zhao, Jiamin~Wu, Tao~Yu$^\dag$ and Hui~Qiao$^\dag$
% <-this % stops a space
\IEEEcompsocitemizethanks{
\IEEEcompsocthanksitem All authors are
from Tsinghua University, Beijing 100084, China. \protect}% <-this % stops an unwanted space
%\thanks{$^\ast$ Co-first author}
\thanks{$^\dag$ Corresponding author}}

% The paper headers
% \markboth{IEEE TRANSACTIONS ON PATTERN RECOGNITION AND MACHINE INTELLIGENCE, Vol. *, No. *, July 2024}{Zhao et al.: PNR: Physics-informed Neural Representation for high-resolution LFM reconstruction}

\IEEEtitleabstractindextext{
\begin{abstract}
Light field microscopy (LFM) has been widely utilized in various fields for its capability to efficiently capture high-resolution 3D scenes.
Despite the rapid advancements in neural representations, there are few methods specifically tailored for microscopic scenes. Existing approaches often do not adequately address issues such as the loss of high-frequency information due to defocus and sample aberration, resulting in suboptimal performance.
In addition, existing methods, including RLD, INR, and supervised U-Net, face challenges such as sensitivity to initial estimates, reliance on extensive labeled data, and low computational efficiency, all of which significantly diminish the practicality in complex biological scenarios.
This paper introduces PNR (Physics-informed Neural Representation), a method for high-resolution LFM reconstruction that significantly enhances performance. Our method incorporates an unsupervised and explicit feature representation approach, resulting in a 6.1 dB improvement in PSNR than RLD. Additionally, our method employs a frequency-based training loss, enabling better recovery of high-frequency details, which leads to a reduction in LPIPS by at least half compared to SOTA methods (1.762 V.S. 3.646 of DINER).
Moreover, PNR integrates a physics-informed aberration correction strategy that optimizes Zernike polynomial parameters during optimization, thereby reducing the information loss caused by aberrations and improving spatial resolution. These advancements make PNR a promising solution for long-term high-resolution biological imaging applications. Our code and dataset will be made publicly available.
\end{abstract}

}

% make the title area
\maketitle

\IEEEdisplaynontitleabstractindextext

\IEEEpeerreviewmaketitle

% \IEEEraisesectionheading{\section{Introduction}\label{sec:introduction}}
\IEEEraisesectionheading{
\section{Introduction}}
\label{sec:introduction}
% \linenumbers
\IEEEPARstart{L}ight field microscopy (LFM), as a classical paradigm, has been used for a long time in observations of biological dynamics and other fields due to its ability to efficiently capture high-resolution 3D microscopic scenes over extended periods\cite{prevedel2014simultaneous,lu2024combining,pegard2016compressive}.
Since the invention of the first light field microscope in 2006\cite{levoy2006light}, the imaging architecture of LFM has been continuously upgraded with a main focus on improving the lateral and axial resolution\cite{wu2021iterative,LFcao,guo2019fourier,rush3d}. 
However, the reconstruction algorithm, RLD-based light field reconstruction method, remains the same.

The advancement of deep learning has offered a promising solution in the microscopic domain for both low-level tasks, such as super-resolution\cite{lu2023virtual,qiao2024zero} and denoising\cite{deepsemi,li2021unsupervised}, as well as high-level tasks, including segmentation\cite{seg} and detection\cite{deepwonder}. 
Although recent deep learning-based microscopic light field reconstruction methods\cite{RLN,vcdnet}, significantly improve the reconstruction efficiency compared with the classical RLD-related methods\cite{laasmaa2011application}, their generalization capacity and reconstruction quality are still far from practical due to the fundamental limitation of the supervised learning paradigm and the lack of real-captured high-resolution training data. As a result, LFM reconstruction with both high resolution and strong generalization is still a challenging open problem. 
Recently, the emergence of implicit neural representation (INR) offers the possibility of high-resolution LFM reconstruction\cite{park2019deepsdf,mildenhall2021nerf,barron2021mip}. INR specifically refers to a class of methods that use neural networks to represent spatial functions. Unlike the traditional discrete explicit representations, by using neural networks to build mappings from spatial coordinates to function values, INR allows for the automated computation of the differentiation of the spatial function, thus offering the possibility of introducing different kinds of constraints during the optimization, or reconstruction, process. 
Methods based on INR have appeared in the LFM field\cite{decaf,DINER}, which have validated the ability to eliminate the missing cone problem and possess strong generalization ability. However, existing INR-based reconstruction methods are too slow (e.g., DECAF needs 20 hours to reconstruct a single sample) to be used for long-term observation. 
Moreover, due to ignoring the lens and sample aberration that prevails in the LFM imaging process, the existing INR-based LFM reconstruction methods do not have obvious resolution advantages, and the comprehensive performance and practicality remain inferior to RLD. 
Even though EFR methods based on 3DGS\cite{3dgs} have surpassed INR-based approaches in macro scenarios, the requirement for precise wave optics modeling in microscopic scenes makes LFM reconstruction using 3DGS exceptionally complex. Therefore, there are no existing methods extending 3DGS to support microscopic light field reconstruction. 

Besides single-frame LFM reconstruction, an efficient and effective sequential LFM reconstruction method is of great practical value since long-term sequential LFM capture and reconstruction is important for applications like in vivo observation. However, the maximum likelihood estimation strategy of RLD methods makes them sensitive to the initial estimate for the iterative optimization, it may fall into local minima even if we use the previous frame as the initial estimate,  this results in an increase in the number of iterations required for reconstruction and deteriorated reconstruction performance. 

To overcome the challenges above and provide a more practical reconstruction solution for improving the spatial resolution of 2pSAM, we propose an Unsupervised and Explicit Feature Representation-based (EFR-based) method for high-resolution LFM reconstruction, named PNR. 
PNR achieves both much higher resolution (4.1dB improvement of PSNR) and efficiency (up to 3x) compared with existing INR-based methods on synthetic data and effectively improves the spatial resolution of 2pSAM on real-world data. 

\begin{figure*}[t]
    \centering
    \includegraphics[width=1\linewidth]{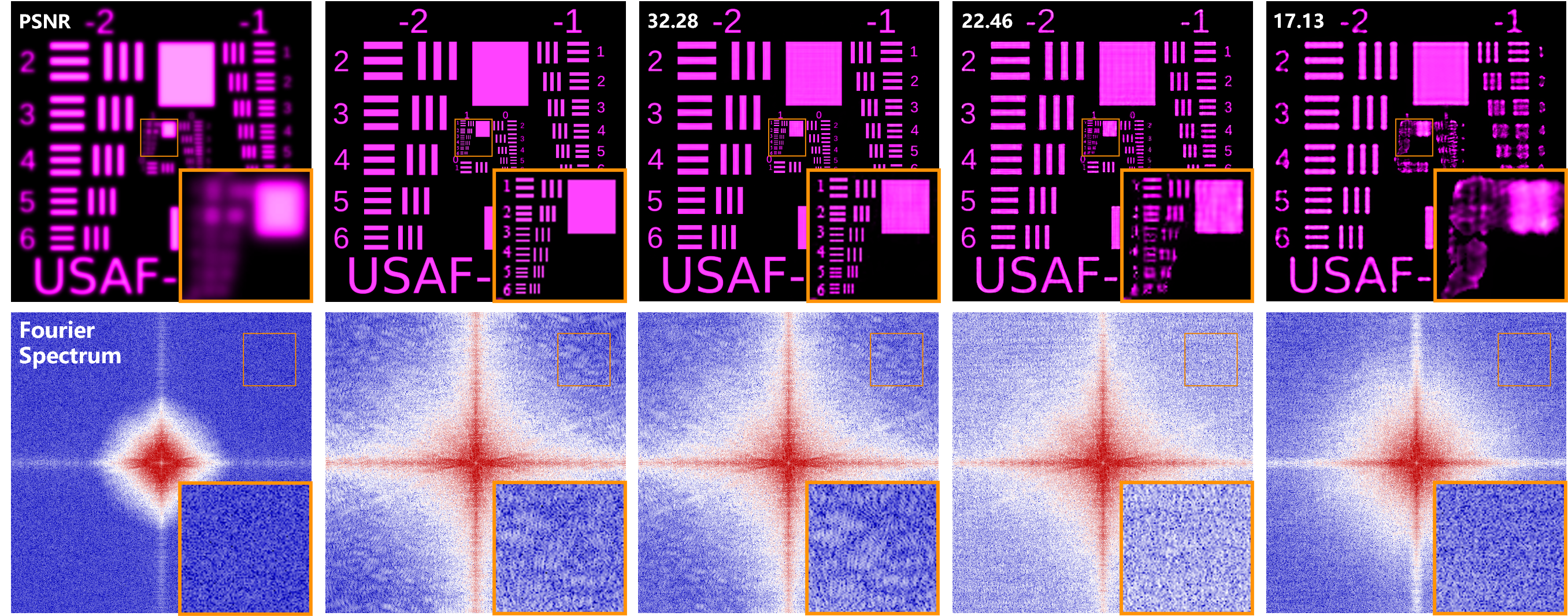}
    \vspace{-0.5cm}
    \begin{center}
\leftline{~LF Image (Center View)\qquad~~Ground Truth\quad\quad\quad~Ours w/ FFT Loss\quad\quad~~Ours w/ SSIM Loss~~~Ours w/ Perception Loss}
    \end{center}  
    \vspace{-0.6cm}
    \caption{Quantitative and qualitative comparison of four loss functions used for image reconstruction (the lower half is the Fourier spectrum of the reconstruction results). 
    %The highlighted area in the center view indicates that spectral cutoff and defocus in LFM can significantly lead to the loss of high-frequency information (the marginal areas of the spectrum).
    The highlighted areas in the spatial and frequency domain of center view indicate that spectral cutoff and defocus in LFM can significantly lead to the loss of high-frequency information.
    Compared to other loss functions, FFT Loss effectively leverages multi-angular information to recover missing high-frequency details.} 
    \label{fig_fftloss_compare}
    \vspace{-0.2cm}
\end{figure*}

The key features of PNR are: 
\begin{itemize}
    \item An efficient EFR for high-resolution LFM reconstruction: Different from the fully implicit mapping based on a single multi-layer perceptron network (which is too compact to reconstruct efficiently), PNR incorporates a dense-feature-based explicit representation with a super-sampling strategy for achieving strong representation ability without sacrificing the reconstruction efficiency. 
    \item A frequency-adaptive training loss: Traditional RLD and INR-based LFM reconstruction methods are more concentrated on recovering low-frequency information, which limits the high-resolution reconstruction capacity and results in a lack of high-frequency reconstruction details (as shown in Fig.\ref{fig_fftloss_compare}). To overcome this limitation, PNR introduces a novel training loss in LFM reconstruction by constructing the loss in the frequency domain directly based on fast Fourier transform (FFT), named FFT loss. The FFT loss acts as an adaptor to balance the training on different frequency bands, which significantly enhances high-frequency information recovery.  
    \item A physics-informed aberration correction strategy: To eliminate the negative impacts caused by aberration and sample diffraction for LFM reconstruction, PNR successfully formulate the impact of aberration according to the physical imaging process and further incorporate the formulation into the reconstruction process. As a result, PNR enables optimizing Zernike polynomial-based aberration parameters directly, which is not only much more efficient (fixed dimension, independent of the input view number) but also much easier to differentiate and optimize. 
    \item The first light field dataset based on two-photon excitation: We utilize 2pSAM as the acquisition device for light field data, as it has been validated to provide high-resolution imaging of deep tissues, particularly in terms of axial resolution\cite{2pSAM}. Due to the use of two-photon excitation, our dataset is less affected by sample scattering compared to other light field datasets, resulting in a higher signal-to-background ratio. Note that this dataset includes not only light field data captured by 2pSAM but also the corresponding 3D intensity volumes obtained by a traditional two-photon microscope as reference images. This dataset will be made publicly available.
\end{itemize}

\section{RELATED WORK}
Light field imaging was initially applied in the realm of natural images\cite{ng2005light}. Due to the ability of light field cameras to simultaneously capture spatial and angular information, they are commonly utilized in tasks such as digital refocusing\cite{refocus,efficientrefocus}, image super-resolution\cite{cheng2021lightSR,wang2022depth}, and depth estimation\cite{opal,OAVC,jin2020deep}. Its application in microscopy began in 2006\cite{levoy2006light}. Since the invention of light field microscopy, its limited spatial resolution has significantly restricted its widespread application\cite{lfreview}. In 2013, Michael Broxton \etal~ introduced concepts from wave optics into the field of LFM, employing wave optics to model the PSF of light field systems and replacing the previous geometrical optics models\cite{broxton2013wave}. At the same time, they applied the Richardson-Lucy deconvolution (RLD) method\cite{fish1995blind} to LFM.
In 2019, Zhi Lu \etal~ proposed a phase space deconvolution method\cite{lu2019phase} for LFM based on RLD method, effectively eliminating artifacts in the 3D reconstruction of linear frequency modulation data and significantly enhancing image contrast and convergence speed. To address the challenges of optical heterogeneity, tissue opacity, and phototoxicity in the in vivo observation, Jiamin Wu \etal~ proposed a computational imaging framework known as DAOSLIMIT\cite{wu2021iterative}. This method employs a unique scanning approach to simultaneously enhance both spatial and angular resolution, achieving ultra-high spatiotemporal resolution in large-scale 3D fluorescence imaging with reduced phototoxicity. 
However, the RLD-based algorithm iteratively corrects the reconstructed volume based on the Poisson assumption, which, while suppressing Poisson noise, can result in excessive smoothing of the images, leading to the loss of some high-frequency details. 
Although more sophisticated RLD-based method have emerged in the field of structured illumination microscopy\cite{zhao2022sparse}, enhancing the resolution of microscopes in some scenarios, their reliance on sparse priors and the use of strong regularization for optimization restrict their applicability to all microscopic contexts.
Additionally, the maximum likelihood estimation strategy employed by RLD method renders them sensitive to the initial estimates used in iterative optimization, making them susceptible to converging on local minima even when the previous frame serves as the initial estimate. This sensitivity results in an increased number of iterations required for reconstruction and deteriorated reconstruction performance.

\begin{figure*}[ht]
    \centering
    \includegraphics[width=1\linewidth]{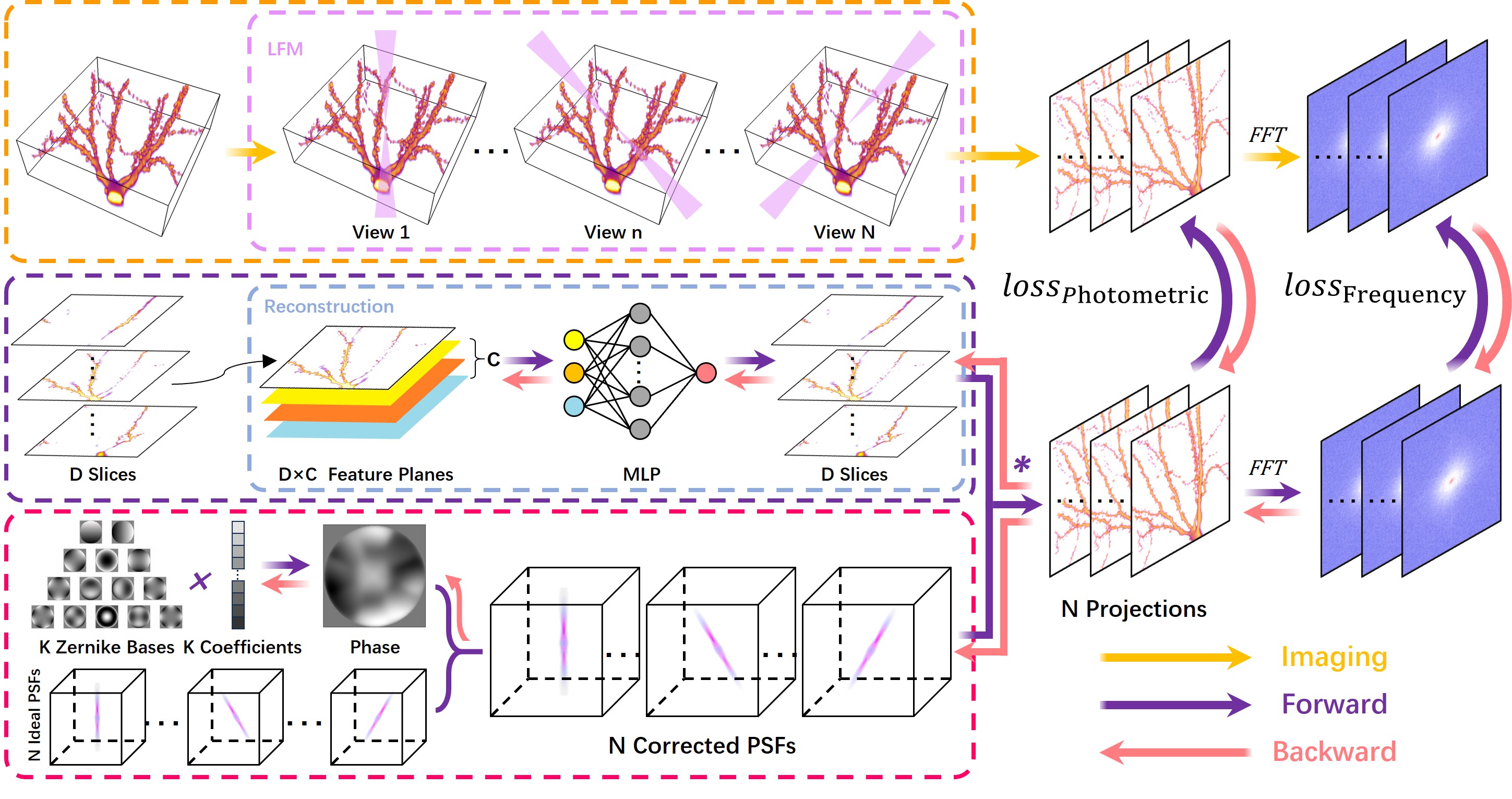}
    \vspace{-0.7cm}
    \caption{Overview of the framework and training loss. Our framework, named PNR, contains four components: explicit feature planes, a two-layer MLP and digital adaptive optics (DAO) module. The yellow dashed box illustrates the actual light field imaging process using 2pSAM.}
    \vspace{-0.4cm}
    \label{fig_framework}
\end{figure*}

Additionally, due to issues such as refractive index inhomogeneity and scattering in the samples, conventional laser microscopy often struggles to achieve near-diffraction-limited performance in deep tissue imaging\cite{zipfel2003nonlinear,cahalan2002two,kong2023neuron}. Due to the longer wavelength (920 nm) of two-photon excitation and its use of localized "nonlinear" excitation, TPM is particularly well-suited for deep tissue imaging\cite{TPM,TPMprinciples}. However, traditional TPM encounters challenges such as phototoxicity and photobleaching at the focal plane\cite{icha2017phototoxicity,patterson2000photobleaching}. Furthermore, due to the nonlinear excitation method employed, TPM is highly sensitive to optical aberrations\cite{hampson2021adaptive}. 
In the field of natural images, employing hardware or specialized chips to mitigate the impact of aberrations on imaging quality has emerged as an effective solution\cite{wu2022aberration,cao2024aberration}.
In 2010, Na Ji \etal~ proposed an adaptive optics method\cite{ji2010adaptive} to enhance the resolution and SNR of optical microscopy, which segments the aperture of the objective lens into multiple sub-aperture regions. By calculating the degree of displacement in each sub-image, the method uses wavefront sensors or spatial light modulators to correct optical aberrations in the microscope. However, it typically necessitates additional wavefront sensors or spatial light modulators to estimate and compensate for wavefront distortions in scattering tissues or imperfect imaging systems. 

Recently, the 2pSAM\cite{2pSAM} has emerged as a novel imaging system that achieves performance approaching the diffraction limit while maintaining low phototoxicity, which is crucial for long-term observations in life sciences. In 2pSAM, a ‘needle’ beam is employed for both 2D spatial scanning and 2D angular scanning, facilitating large-field 3D high-speed imaging with subcellular resolution. This is also accomplished through the application of a RLD-based reconstruction algorithm which still limits the further enhancement of the resolution of 2pSAM.

Although recent deep learning-based microscopic light field reconstruction methods\cite{RLN,vcdnet} significantly improve the reconstruction efficiency compared with the classical RLD-related methods, their generalization capacity and reconstruction quality are still far from practical due to the fundamental limitation of the supervised learning paradigm and the lack of real-captured high-resolution training data. As a result, LFM reconstruction with both high resolution and strong generalization is still a challenging open problem\cite{verinaz2021deep}. 
Since 2020, implicit neural representation (INR) becomes a hot tool in the computer vision and graphics community for its superior performance on tasks like novel view synthesis\cite{ying2023parf,mildenhall2021nerf,barron2021mip}, 3D reconstruction\cite{wang2021neus,liu2024finer,pamir} and even physical simulation\cite{qiao2022neuphysics,wanghx}. 
The vanilla INR uses a single coordinate-based MLP to represent the target vector field. Compared with traditional discrete representations like Multi-plane-image (MPI) and cost volumes\cite{he2024mmpi,costvolume}, INR achieves much higher reconstruction quality and stability. Additionally, by using neural networks to build mappings from spatial coordinates to function values, INR allows for the automated computation of the differentiation of the spatial function, thus offering the possibility of introducing different kinds of constraints during the optimization, or reconstruction, process. 
Methods based on INR have appeared in the LFM field\cite{decaf}, which have validated the ability to eliminate the missing cone problem and possess strong generalization ability. However, existing INR-based reconstruction methods are too slow (e.g., DECAF needs 20 hours to reconstruct a single sample) to be used for long-term observation. However, the single-MLP representation is too compact which leads to: i) a significant decrease on the efficiency and ii) severe artifacts like floaters which is unacceptable when the 3D structure is the main consideration. 
DINER\cite{DINER} significantly enhances reconstruction speed while improving accuracy. However, it still fails to effectively mitigate the loss of high-frequency information caused by the spectral cutoff and defocus in microscopic scenes. Additionally, the use of the SIREN function contributes to instability in the reconstruction.

Recently, methods based on explicit feature representation (EFR) have demonstrated performance comparable to state-of-the-art implicit representation techniques in the realm of macro-scale 3D reconstruction, while significantly surpassing them in terms of efficiency (e.g., Instant-NGP\cite{muller2022instant}, TensoRF\cite{chen2022tensorf} and 3DGS\cite{3dgs}). However, the effectiveness of EFR-based methods has yet to be validated in the microscopic domain.
Moreover, due to ignoring the optical and sample aberration that prevails in the imaging process, the existing INR-based LFM reconstruction methods do not have obvious resolution advantages, and the comprehensive performance and practicality remain inferior to RLD. 

\begin{figure*}[ht]
    \centering
    \includegraphics[width=1\linewidth]{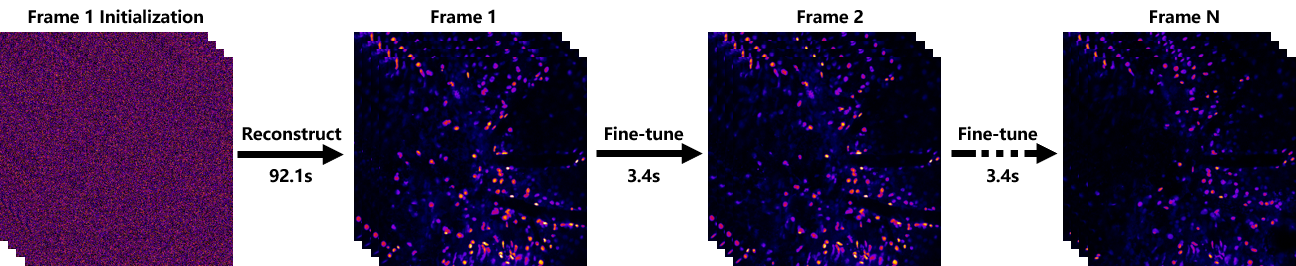}
    \vspace{-0.4cm}
    \caption{Schematic diagram of sequential reconstruction. The sequential reconstruction process begins with a random initialization of Frame 1, followed by reconstruction (92.1s) and fine-tuning of subsequent frames (3.4s). Given the similarity between adjacent frames, it is feasible to  expedite dynamic reconstruction by explicitly fine-tuning the feature vectors.}
    \vspace{-0.4cm}
    \label{fig_dynamic_sketch}
\end{figure*}

\section{METHOD}
In this section, we will detail the methodology for high-resolution light field microscopy reconstruction. We will begin with a brief introduction to the 3D Imaging based on LFM, emphasizing its capabilities and significance.  Next, we will discuss the overall framework process, highlighting the significance of explicit feature representation and our training strategy. We will then introduce the approach for aberration estimation to ensure accuracy in the reconstruction process. Furthermore, we will delve into the design of the FFT loss function, explaining its critical role in the optimization process. Finally, we will supplement additional details to enhance the reader's understanding of the method.
\subsection{3D Imaging based on LFM}
%The 2pSAM system operates by utilizing a unique ‘needle’ beam for advanced light field imaging. This system enables both two-dimensional (2D) spatial scanning and 2D angular scanning, which together facilitate large-field three-dimensional (3D) imaging at high speeds and with subcellular resolution. Depth-of-field expansion is achieved through low numerical aperture (NA) excitation, with the pinhole positioned at the conjugate plane of the imaging plane, ensuring its diameter is aligned with the diffraction limit of low-NA excitation. This configuration transforms the Gaussian beam into an Airy disk-like special beam, effectively preserving high-frequency components during high NA objective imaging.
Light field microscopy employs two-dimensional angular scanning techniques, such as LED multi-angle illumination and microlens arrays, to achieve three-dimensional (3D) imaging with high speed and subcellular resolution.
To analog the real-world imaging system, we formulate the influence from the optical lens to the sample. First of all, we derive the optical transfer function (OTF) representation by modeling the light propagation process in a wave optics framework, encompassing the journey from the laser output to the objective plane. Then, We denote the direction along light as $Z$ axis and sample $z$ points, the intensity of each point is denoted as $I_{x,y,z}$. As shown in Fig.\ref{fig_framework}, there are $U$ beams illuminating the sample from different angles, each modeled as a optical transfer function, denoted as $OTF_{u,z}$. Correspondingly, each point spread function (PSF) can be represented as
\begin{equation}
PSF_{u,z} = \|FFT(OTF_{u,z})\|_2^4,
\end{equation}
and the observed projection is denoted as
\begin{equation}
LF_{u,x,y} = \sum_{z}(I_{x,y,z} * PSF_{u,z}),
\end{equation}
where $FFT()$ indicates the fast fourier transformation and $*$ means 2D convolution operation. 
%Generally, we use 13 or 35 angles for 3D reconstruction.

\begin{figure*}[p]
    \centering
    \includegraphics[width=1\linewidth]{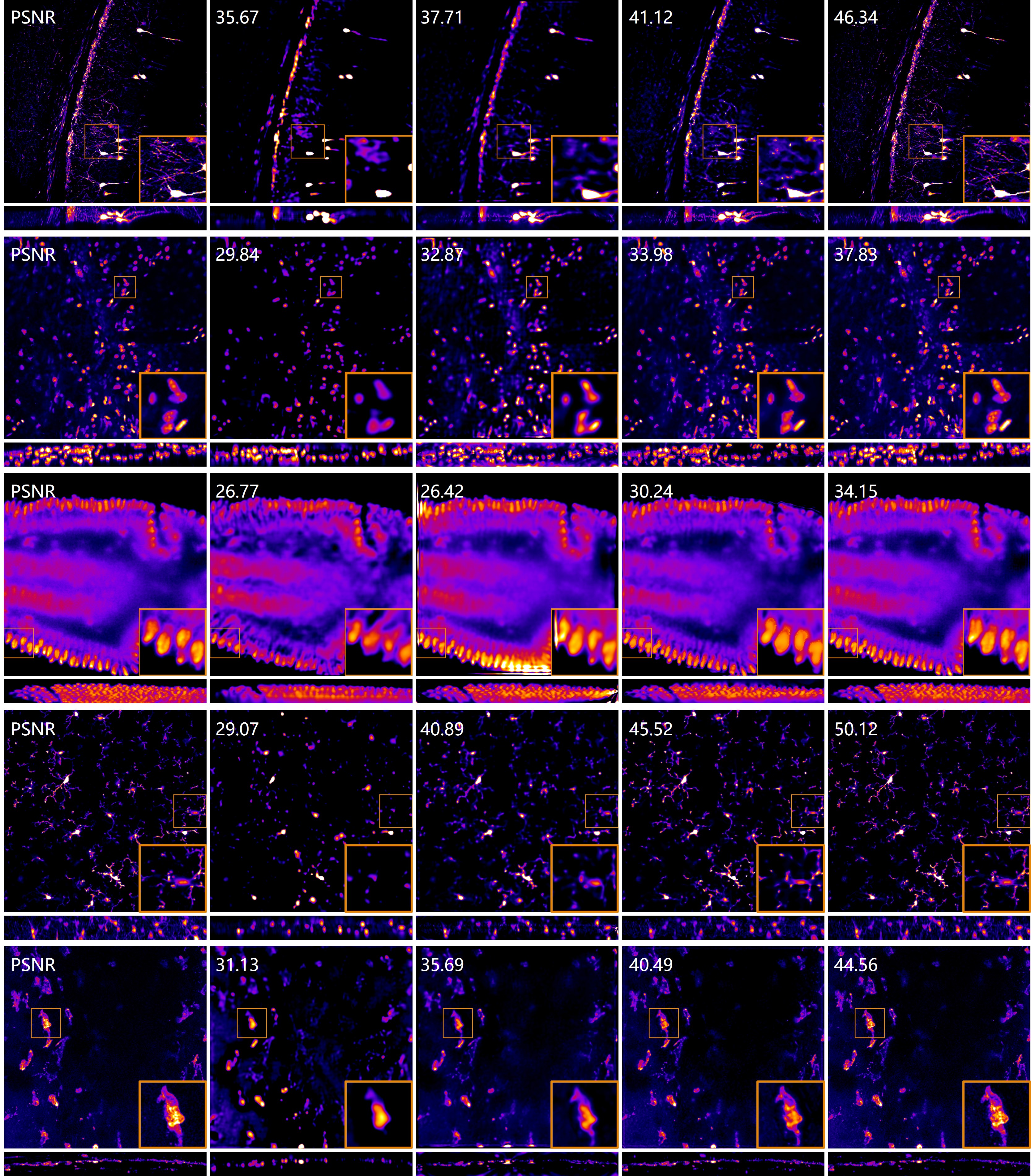}
     \begin{center}
     \vspace{-0.2cm}
     \leftline{\qquad\qquad\quad~GT\qquad\qquad\qquad\quad~~~VCD-Net~~~\quad~\qquad\qquad~~~~~~RLD~~~~~~\qquad\qquad\qquad~~DINER~~\qquad\qquad\qquad\quad~Ours}
  \end{center}  
  \vspace{-0.6cm}
  \caption{Qualitative and quantitative comparison of SOTA methods on the synthetic dataset. Five biological samples arranged from top to bottom are mouse neurons, immune cells, drosophila embryo, microglia in mice and microglia in mice after traumatic brain injury (TBI). Note that due to the presence of large areas without samples in fluorescent microscopy images, the computed PSNR values are often higher than those for natural images.}
  \label{fig_syn_compare}
\end{figure*}

\subsection{End2End Optimization Framework.}

In this section, we analyze in-depth the specific designs in our LFM reconstruction framework contribute to performance and efficiency improvement.

\subsubsection{Digital Adaptive Optics}
Inevitably, manufacturing errors of optical elements, inhomogeneous sample and scattering effects can introduce aberrations in the optical system. One of the benefits of our reconstruction framework is that one can easily incorporate complicated formulations or constraints into the forward imaging process for LFM reconstruction without explicitly differentiating them. 

Since aberration is inevitable and sample aberration calibration or estimation is an additional step which relies on expensive hardwares (wavefront sensors or spatial light modulators) and complicated operations in traditional adaptive optics. As shown in the bottom left section of Fig.\ref{fig_framework}, we formulate the impacts of aberration using Zernike Polynomial and incorporate this into the forward imaging module and achieves End2End phase optimization together with the LFM reconstruction, named digital adaptive optics (DAO). In other words, aberration describes the degree of imperfection in the imaging of optical systems. The Zernike coefficient\cite{zernike}, which is consistent with the polynomial form of the same aberration, is used to fit the wavefront and describe the optical aberration.
The Zernike polynomial can decompose wavefront aberrations into higher-order components and each Zernike basis is denoted as $Z_k$, where $k$ indicates the decomposition level. For each Zernike basis, we assign a learnable parameter $P_k$ as the weight to jointly affect the aberration. Thus, the aberration is denoted as
\begin{equation}
Abe = \sum_k Z_k\cdot P_k.
\end{equation}
Then, the practical PSF can be represented as 
\begin{equation}
PSF_{u,z} = \|FFT(OTF_{u,z}\cdot e^{-i\cdot Abe})\|_2^4.
\end{equation}

Due to the influence of aberrations on forward projections from all angles, we can leverage this constraint to simultaneously perform volume reconstruction and aberration correction during the end-to-end optimization process. 

\subsubsection{Explicit Feature Representation for Intensity Field}

While the MLP continuously represents the 3D scene by taking any position in the volume as an input, the fully implicit MLP function makes it difficult to achieve full exploitation of resolution for 3D scene representation. Specifically, since implicit representations maintain 3D consistency, observed pixels absorb information from nearby views, which inevitably leads to a decrease in resolution.
Different from the vanilla INR using a single coordinate-based MLP , we propose to use explicit feature representation to describe the intensity of the observed microscopic samples. 

As shown in the middle part on the left side of Fig.\ref{fig_framework}, we establish an explicit three-dimensional volume $V$ including $C$ feature planes, each position $x,y,z$ describes the feature of the sample with the same position. 
%The explicit feature representation is a 3D feature volume together with a 2-layer MLP network. The 3D feature volume covers the target reconstruction space and was uniformly discritized into voxels, each voxel stores a learnable feature vector. The target intensity volume was acquired by simply mapping each feature vector to the intensity value through the light-weight 2-layer MLP network. (Fig.\ref{fig_framework}). 
%As shown in the middle part on the left side of Fig.\ref{fig_framework}, explicit feature representation decompose the single large MLP into $C$ explicit feature planes together with a light-weight 2-layer MLP. Note that each point of the sample corresponds to a feature vector with $C$ channels.
Note that each point of the sample corresponds to a feature vector with $C$ channels. Later, we introduce a light-weight 2-layer MLP, which considers the explicit feature as input and generates the final intensity value. We define the MLP network as an mapping function $f$, and the output is denoted as 
\begin{equation}
\label{q1}
I_{x,y,z} = f(V_{x,y,z}).
\end{equation}

Both the explicit feature volume and the MLP are learnable and trained concurrently. 
The explicit feature planes are fully discritized so that each feature vector can be optimized independently in parallel to improve the reconstruction efficiency and eliminate floater artifacts. 
Moreover, the light-weight 2-layer MLP is only responsible for mapping but not reconstructing to ease the optimization and produce high quality reconstruction.  

\subsubsection{Sequential Reconstruction}
Different from RLD methods, we find that the explicit feature representation together with the auto differentiation and gradient descent optimizer is more robust against the initial estimate and thus more suitable for directly fine-tuning the previous frame for the reconstruction of the current frame. 

As shown in Fig.\ref{fig_dynamic_sketch}, the reconstructed results of the previous frame can be fully utilized as the initial estimate to accelerate reconstructing the current frame. In this way, after conducting long-term imaging of biological samples, it is feasible to spend a relatively long time on the initial reconstruction of the first frame, and then sequentially reconstruct each subsequent frame based on that to realize efficient sequential reconstruction. Generally, this approach can achieve a speedup of over 20 times. As shown in Tab.\ref{tab_dynamic}, the overall framework of our method enables an additional 2X reconstruction speed improvement for sequential reconstruction compared with RLD method. 

\subsection{Training Losses}
One of the main challenges of LFM reconstruction is recovering high-frequency and detailed 3D structures. 
Existing unsupervised learning-based LFM reconstruction methods mainly rely on the intensity-based MSE loss for training. 
Although this loss is effective enough for light field macroscopy reconstruction, the performance drops a lot for light field microscopy reconstruction since the optical defocus will result in significant blurry images with high-frequency details over-smoothed. 
Therefore, we propose to additionally map the projections into the frequency domain for loss construction. 
We also designed two regularization losses to reduce outliers in the reconstruction results. 
\subsubsection{FFT Loss}
Since the MSE or L1 loss reduces the sharp details because of the intrinsic attributes of  $L_1$ structure, which optimizes the average value of multiple pixels. We introduce the FFT loss for detail generation. As shown in Fig.\ref{fig_framework}, by mapping the images using the FFT, the FFT loss successfully balanced the optimization of information in different frequencies: taking more attention to the reconstruction of high-frequency details without sacrificing the reconstruction accuracy of the principle (low-frequency) components.
Specifically, we calculate the mean squared error between the generated and the real-captured LF images in both the spatial domain and the Fourier domain for optimization. We define the $LF_{i}$ as the ground-truth value of real-captured projectio pixel $i$. The corresponding value of simulated pixel is denoted as $\hat{LF}_i$. The MSE loss is defined as 
\begin{equation}
L_{MSE} = \sum_{i}(LF_{i} - \hat{LF}_i )^2,
\end{equation}
and the FFT loss is indicated as
\begin{equation}
L_{FFT} = \frac{\sum_{i}(FFT(LF_{i}) - FFT(\hat{LF}_i) )^2}{N},
\end{equation}
where $N$ is the total number of the pixels.

As shown in Fig.\ref{fig_fftloss_compare}, the FFT loss could significantly improves the high-frequency reconstruction performance compared to other loss functions used for image super-resolution. And the lower part of Fig.\ref{fig_fftloss_compare} demonstrates that the spectrum of the reconstruction results using FFT loss is more complete. As a result shown in Tab.\ref{tab_ablation}, the incorporation of the FFT Loss achieves a 3X LPIPS\cite{lpips} decrease (from 1.533 to 0.468).

\subsubsection{Regularization}
First, to minimize the 'floaters' caused by noise in the reconstruction results, we applied a total variation loss function as a a regularization term to penalize the variation along Z-axis. During training, the Z-TV loss $L_{Z-TV}$ is presented as
\begin{equation}
L_{Z-TV} = \sum_{Z}|\hat{I}_{x,y,z} - \hat{I}_{x,y,z-1}|.
\end{equation}

Since many areas of the scene lack samples or have very low sample intensity, negative values can easily arise during the optimization process. Using the ReLU activation function may cause some areas to remain inactive, while using the leaky ReLU activation function\cite{leackyrelu} can still lead to negative values. Therefore, we apply the positive loss to penalize the occurrence of negative values while using the leaky ReLU activation function. Specifically, the positive loss is indicated as
\begin{equation}
L_{pos} = \sum_{x,y,z}ReLU(-\hat{I}(x,y,z)).
\end{equation}

%In summary, we digitally modeling the physical imaging process of the scanning light field microscopy system we used, which is shown in Fig.~\ref{framework}a. 
%Both lens aberration and sample aberration affect the imaging quality. 
%As shown in Fig.~\ref{framework}b, we first establish the PSF simulation lens aberration generation based on a priori knowledge. 
%{Subsequently, we construct the Zernike polynomial fitting, which is considered as the representation of the sample aberration, and convolve it with the lens aberration to realize the simulation in imaging loss (top of Fig.~\ref{framework}c).} 
%The 3D samples are described by hashed-feature-based implicit representations. (no hash)
%Sample points in 3D space are represented uniformly by an MLP(bottom of Fig.~\ref{framework}c). 
%The lossless projection of the samples is convolved with the aberration to generate the output of the light field. }
\begin{table}[ht]
    \centering
    \caption{Quantitative comparison (PSNR, SSIM and LPIPS) with state-of-the-art methods on synthetic dataset. The best results are in red.}
    \begin{tabularx}{0.45\textwidth}{@{}l*{5}{>{\centering\arraybackslash}X}@{}}
        \toprule
        Scene & Metric & VCD-Net & RLD & DINER & Ours \\ \midrule
        Neurons & PSNR & 46.13 & 47.73 & 48.66 & \textcolor{red}{52.54} \\
          in drosophila & SSIM & 0.927 & 0.983 & 0.986 & \textcolor{red}{0.989} \\
          & LPIPS & 1.171 & 1.261 & 1.679 & \textcolor{red}{0.403} \\ \midrule
        Microglia & PSNR & 36.84 & 35.96 & 36.82 & \textcolor{red}{40.57} \\
          in mice after & SSIM & 0.793 & 0.903 & 0.939 & \textcolor{red}{0.948} \\
          TBI & LPIPS & 7.703 & 6.111 & 4.153 & \textcolor{red}{2.454} \\ \midrule
        Drosophila & PSNR & 27.84 & 20.31 & 26.71 & \textcolor{red}{30.53} \\
         embryo & SSIM & 0.839 & 0.829 & 0.851 & \textcolor{red}{0.923} \\
          & LPIPS & 10.91 & 11.66 & 10.01 & \textcolor{red}{5.175} \\ \midrule
        Microglia & PSNR & 42.03 & 43.92 & 45.58 & \textcolor{red}{50.48} \\
          in mice & SSIM & 0.921 & 0.972 & 0.981 & \textcolor{red}{0.992} \\
          & LPIPS & 7.823 & 6.273 & 3.556 & \textcolor{red}{1.111} \\ \midrule
        Neurons & PSNR & 37.11 & 38.44 & 39.33 & \textcolor{red}{45.45} \\
          in mice & SSIM & 0.791 & 0.916 & 0.929 & \textcolor{red}{0.941} \\
          & LPIPS & 12.54 & 11.07 & 5.721 & \textcolor{red}{3.063} \\ \midrule
        Neurons & PSNR & 44.12 & 46.99 & 48.27 & \textcolor{red}{50.53} \\
          in mice & SSIM & 0.911 & 0.954 & 0.966 & \textcolor{red}{0.968} \\
          & LPIPS & 3.333 & 2.682 & 2.052 & \textcolor{red}{0.417} \\ \midrule
        Vessel & PSNR & 41.43 & 45.44 & 47.76 & \textcolor{red}{52.19} \\
          & SSIM & 0.744 & 0.968 & 0.972 & \textcolor{red}{0.992} \\
          & LPIPS & 1.771 & 1.529 & 1.516 & \textcolor{red}{0.474} \\ \midrule
        Dendrites & PSNR & 35.98 & 37.78 & 38.55 & \textcolor{red}{42.68} \\
          of neurons & SSIM & 0.726 & 0.902 & 0.922 & \textcolor{red}{0.934} \\
          & LPIPS & 8.171 & 5.429 & 2.569 & \textcolor{red}{1.561} \\ \midrule
        Vessel & PSNR & 44.23 & 47.43 & 48.48 & \textcolor{red}{54.73} \\
          & SSIM & 0.821 & 0.971 & 0.989 & \textcolor{red}{0.994} \\
          & LPIPS & 1.901 & 1.584 & 1.411 & \textcolor{red}{0.757} \\ \midrule
        Immune cells & PSNR & 29.42 & 30.81 & 35.12 & \textcolor{red}{39.17} \\
           & SSIM & 0.476 & 0.823 & 0.905 & \textcolor{red}{0.935} \\
           & LPIPS & 11.08 & 9.919 & 3.801 & \textcolor{red}{2.211} \\ \midrule
        Average & PSNR & 38.51 & 39.48 & 41.53 & \textcolor{red}{45.59} \\
             & SSIM & 0.794 & 0.922 & 0.944 & \textcolor{red}{0.962} \\
             & LPIPS & 6.64 & 5.751 & 3.646 & \textcolor{red}{1.762} \\ \bottomrule
    \end{tabularx}
    \vspace{-0.5cm}
    \label{tab_synthetic}
\end{table}

\begin{figure}[t]
    \centering
    \includegraphics[width=1\linewidth]{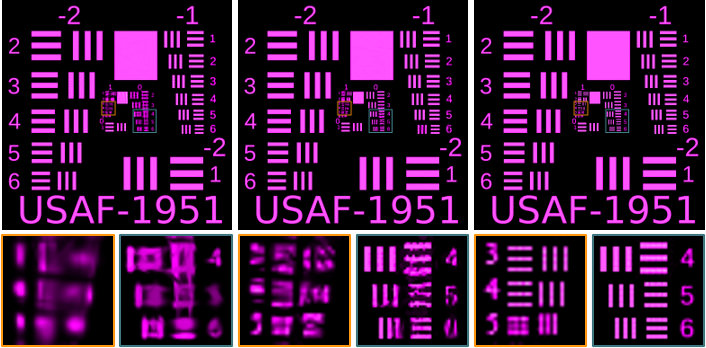}
    \vspace{-0.6cm}
    \begin{center}
    \leftline{\qquad~~~3DGS~\quad\quad\quad~~3DGS w/ FFT loss~~\qquad\quad~Ours}
  \end{center} 
   \vspace{-0.8cm}
  \caption{Qualitative comparisons of 3DGS and our method on USAF resolution test chart. As shown in the highlighted regions, FFT Loss helps the 3DGS reconstruct more high-frequency details, but the overall quality is still inferior to ours.}
    \label{fig_3dgs_compare}
  \vspace{-0.4cm}
\end{figure}

\subsection{Network Architecture and Implementation Details}

\begin{figure*}[ht]
    \centering
    \includegraphics[width=1\linewidth]{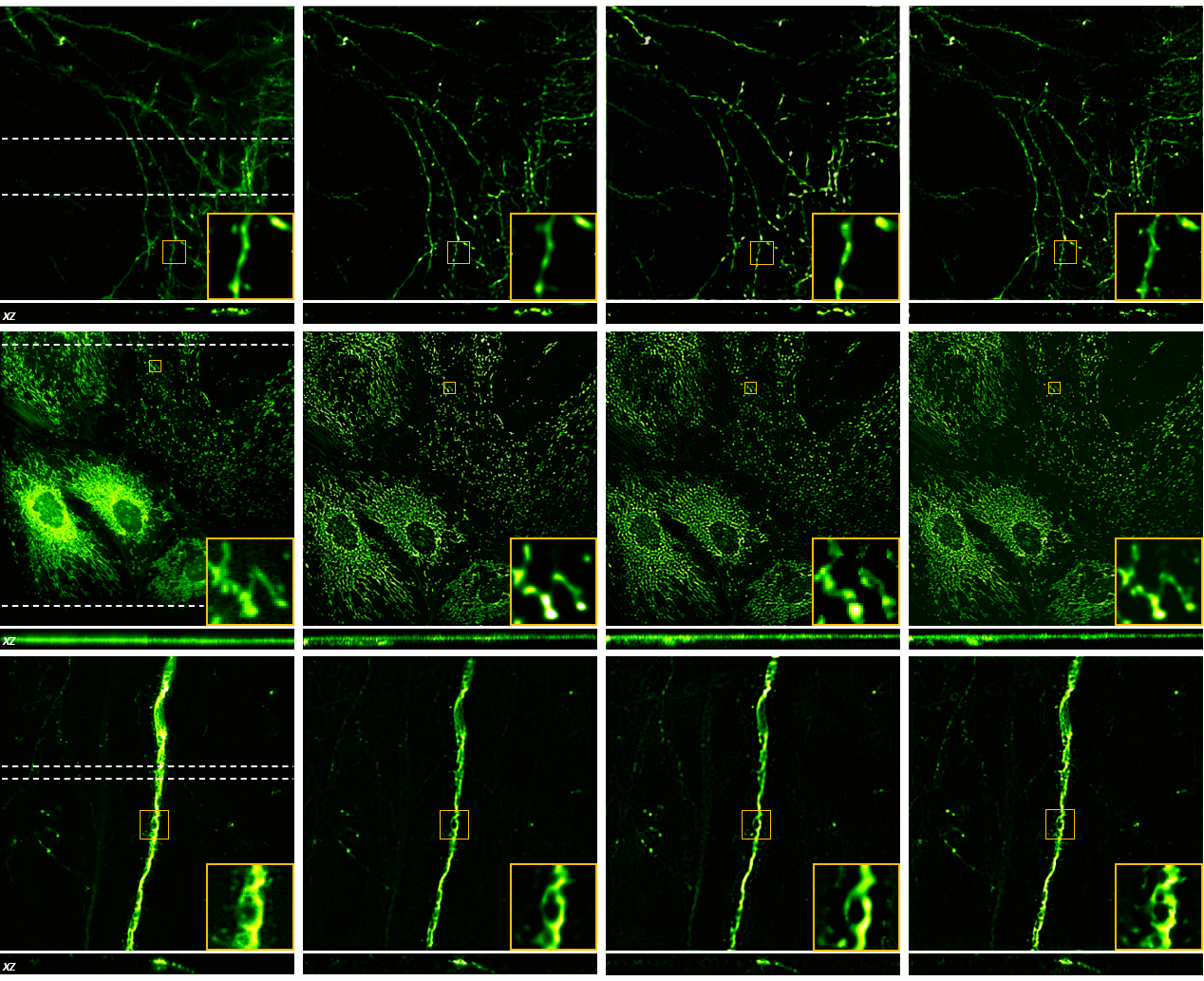}
     \begin{center}
     \vspace{-0.2cm}
    	\leftline{\qquad\qquad\qquad~GT\qquad\qquad\qquad\qquad\qquad\quad~~~RLD\qquad\qquad\qquad\qquad\qquad~~~DINER~~\qquad\qquad\qquad\qquad\qquad~~Ours}
  \end{center}  
  \vspace{-0.6cm}
  \caption{Qualitative comparison of SOTA methods on the synthetic dataset. Three biological samples arranged from top to bottom are dendrites of neurons, endoplasmic reticulum, and nerve fibres. The white dashed lines indicate the range of the computed axial projections.}
  \label{fig_real_compare}
  \vspace{-0.2cm}
\end{figure*}

As shown in Fig.\ref{fig_framework}, the input of our method is the real-captured multi-view 2D images of a physical sample from the 2pSAM. The output is the high-resolution 3D intensity volume of the physical sample. 
The overall light field microscopy reconstruction framework consists of: 
i) an explicit feature representation of the target 3D intensity volume (Figure.\ref{fig_framework}) ii) a forward imaging module consumes the feature representation and produces multi-view 2D projections by modeling the imaging process of the light-field microscopy system (Figure.\ref{fig_framework}) iii) various loss functions between the multi-view 2D projections and the real-captured multi-view 2D images (Figure.\ref{fig_framework}) for optimizing the parameters in i) and ii) simultaneously through iterative loss backward propagation. Based on the explicit feature representation and the forward imaging module, we digitized the whole physical imaging process in which the parameters that need to be optimized during the reconstruction process are: feature vectors in the 3D feature volume, weights of the 2-layer MLP and coefficients of the Zernike polynomial. All these parameters are optimized simultaneously by iteratively and automatically back-propagating the training losses between the simulated projections and the real-captured LF images. 

We employed a robust computational setup to ensure efficient processing and accurate results. In the digital adaptive optics module, we used up to the 45th order Zernike polynomial to represent wavefront aberrations, which offered a balance between computational complexity and accuracy in representing the optical aberrations. The loss weight in our optimization framework is assigned a specific value to balance their contributions effectively. The training losses are composed of four main loss functions, i.e., the MSE loss $L_{MSE}$, the FFT loss $L_{FFT}$, the Z-TV loss $L_{Z-TV}$, and the positive loss $L_{pos}$.
Thus, the final training loss $L_F$ is denoted as 
\begin{equation}
L_F = L_{MSE} + \alpha L_{FFT} + \beta L_{Z-TV} + \gamma L_{pos}.
\end{equation}
where $\alpha$, $\beta$ and $\gamma$ indicate the weights of loss functions. Specifically, we set the weight $\alpha=1e-3$, $\beta=1e-2$ and $\gamma=1e-2$.

Furthermore, we use super-sampling\cite{ssp} to improve the detail quality of the reconstructed 3D volume. Given the 3D volume size as $X\times Y\times Z$, we determine the scale factor as $s$ and the sampling resolution as $sX\times sY\times sZ$. In this case, each point of the scene is divided into $s^3$ sub-points. During the reconstruction process, we use a feature volume with a resolution of $sX\times sY\times sZ\times C$ to represent the entire scene. After mapping it through the MLP to generate an intensity volume, we then downsample the intensity volume to $X\times Y\times Z$ resolution and convolve it with PSFs to produce 2D projections at a resolution of $X\times Y$, which is identical to that of the real-captured LFs.

For the hardware configuration, we utilized an NVIDIA GPU 3090 to accelerate our computations, which provided the necessary computational power for handling the large-scale data and complex calculations involved in our study. 

\begin{figure}[h]
    \centering
    \includegraphics[width=1\linewidth]{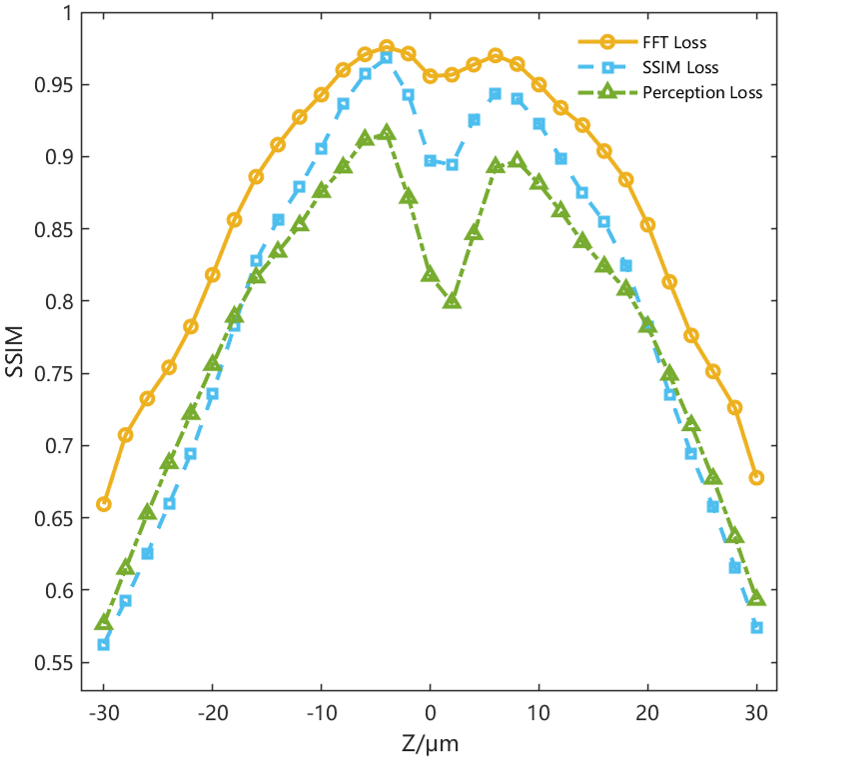}
    \vspace{-0.6cm}
    \caption{Quantitative comparison of reconstructions for each slice along the Z-axis of the focal stack (mouse neuron) using three loss functions. The focal stack consists of 61 slices, with $Z$ representing the distance between the reconstructed slices and the focal plane ($Z = 0\mu m $) of the 2pSAM.}
    \vspace{-0.4cm}
    \label{fig_loss_plot}
\end{figure}

\section{EXPERIMENTAL RESULTS}
In this section, we first introduce the datasets and metrics used for evaluation. Then we report the quantitative and qualitative comparisons with state-of-the-art methods. Finally, we perform ablation studies to analyze different components of our method.

\subsection{Datasets and Evaluation Criteria}
We verified the superior performance of PNR using both simulation data and real data. We capture real-world data using the 2pSAM. 
As an unsupervised optimization reconstruction method, does our method significantly improve the overall performance of existing learning-based methods. 

\begin{figure}[ht]
    \centering
    \includegraphics[width=1\linewidth]{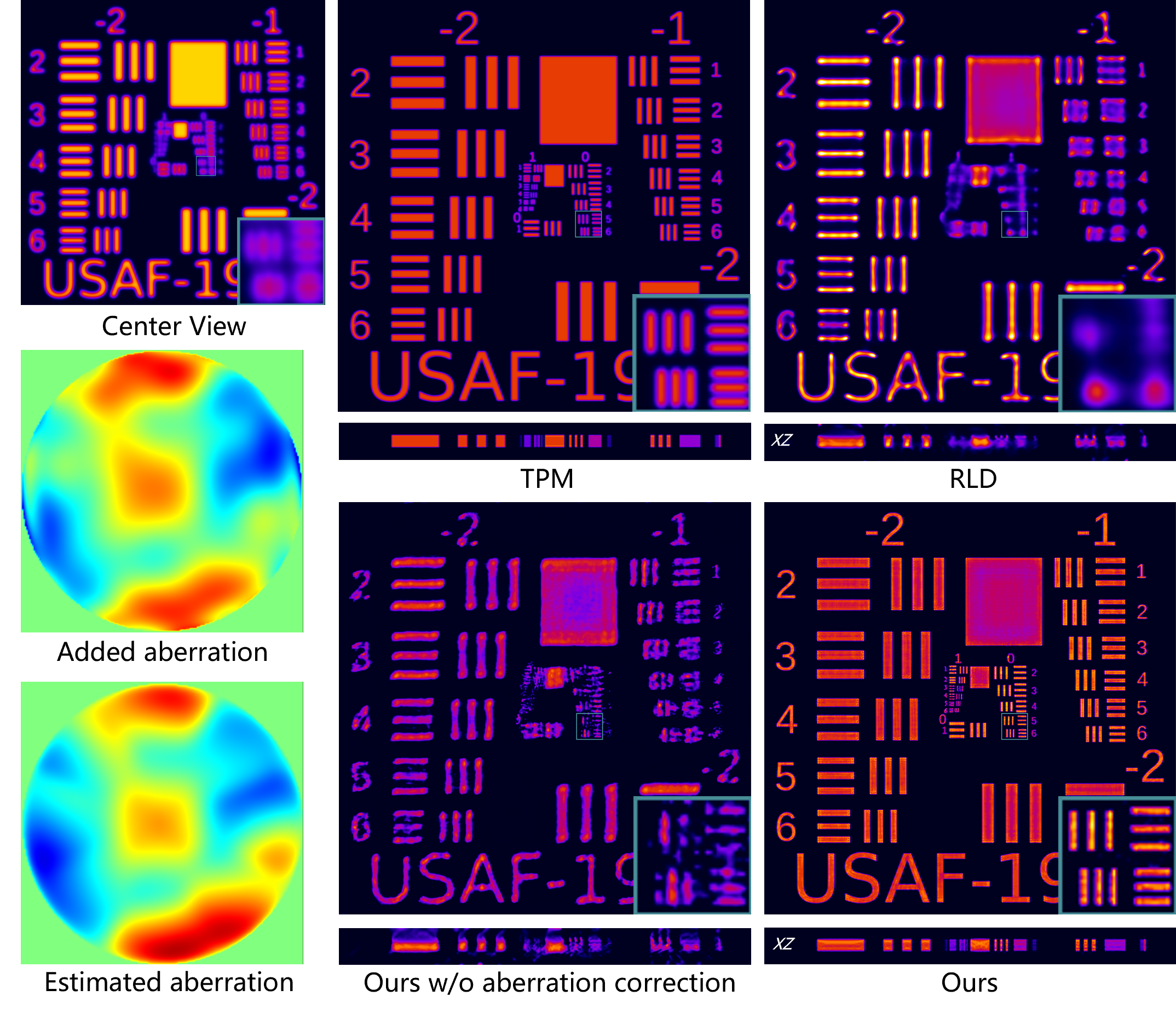}
   \vspace{-0.6cm}
   \caption{Validation of our aberration correction method on the USAF resolution test chart. The "TPM" represents simulations performed using a traditional two-photon microscope without added aberrations, while the "Center View" is obtained using 2pSAM, with added aberrations introduced at the pupil plane during the simulation.}
   %\vspace{-0.4cm}
   \label{fig_syn_DAO}
\end{figure}

We adopt the 2pSAM scans to achieve optical sampling of observing 3D volume, as shown in the top of Fig.\ref{fig_framework}.
The 2pSAM system operates by utilizing a unique ‘needle’ beam for advanced light field imaging. This system enables both two-dimensional (2D) spatial scanning and 2D angular scanning, which together facilitate large-field three-dimensional (3D) imaging at high speeds and with subcellular resolution. Depth-of-field expansion is achieved through low numerical aperture (NA) excitation, with the pinhole positioned at the conjugate plane of the imaging plane, ensuring its diameter is aligned with the diffraction limit of low-NA excitation. This configuration transforms the Gaussian beam into an Airy disk-like special beam, effectively preserving high-frequency components during high NA objective imaging.
Through the rotation of the mechanism, we capture 13 or 35 projections of different angles on the 3D sample, each of which represents the observation in a different view direction. All the datasets captured by the 2pSAM will be made publicly available soon.

\begin{figure*}[h]
    \centering
    \hspace{-0.2cm}\includegraphics[width=1\linewidth]{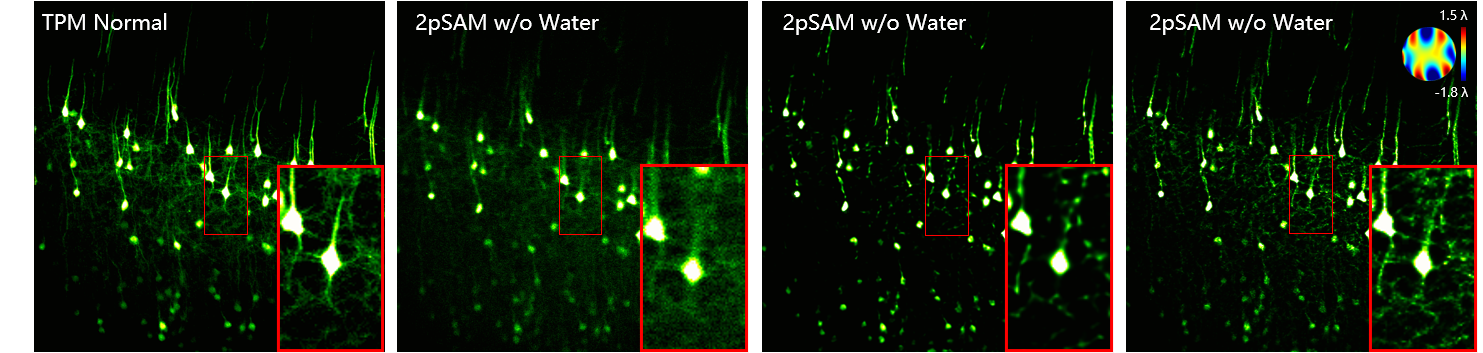}
     \begin{center}
     \vspace{-0.2cm}
     \leftline{\qquad\qquad\qquad~~~GT\qquad\qquad\qquad\qquad\quad~~~Center View\qquad\qquad\qquad\quad~Ours (DAO off)~~\qquad\qquad\quad~~Ours (DAO on)}
  \end{center}  
  \vspace{-0.7cm}
  \caption{Validation of digital adaptive optics (DAO) on real-world data (mouse neurons).}
  \label{fig_real_DAO}
  \vspace{-0.2cm}
\end{figure*}

For the synthetic dataset, we used a traditional high-resolution two-photon microscope ($NA=1.05, \lambda=920nm$) to obtain 3D intensity maps as ground truth. We observed 10 biological samples including mouse neurons, fly embryo, and so on. Notably, we conducted a long-term observation of the microglia cells (1000 frames).
Through denoising and resizing operations, we obtained 10 high-SNR 3D intensity maps of the biological samples, each with a resolution of $512 \times 512 \times 6$. Additionally, we used the USAF resolution test chart\cite{zhangyi} for qualitative comparison. Subsequently, we used the PSFs of 2pSAM and the 3D intensity maps to perform physics-based forward projection, resulting in simulated light field images.

For the real-world dataset, we used the same traditional two-photon microscope to obtain high-resolution 3D intensity maps. We observed 6 kinds of biological samples including neurons, Endoplasmic Reticulum(ER), and so on. Then we used 2pSAM for multi-angle observation to obtain light field images. For low-SNR light field images, we employed DeepCAD\cite{deepcad,deepcadrt} for denoise and NoRMCorre\cite{pnevmatikakis2017normcorre} for rigid motion correction. Additionally, to further demonstrate the aberration robustness of our method, we imaged mouse brain slice with aberrations caused by a lack of water between the objective and the sample.

For each method, we uesd PSNR, SSIM, and LPIPS to evaluate the accuracy of the reconstruction. Note that We defined the LPIPS as
\begin{equation}
LPIPS = \sum_{Z}||f_{Alex}(I_{x,y,z}) - f_{Alex}(\hat{I}_{x,y,z})||_2,
\end{equation}
where $f_{Alex}$ is the pre-trained AlexNet.

\subsection{Comparisons with State-of-the-art Methods}
\noindent\textbf{Methods for comparison.} 
We quantitatively and qualitatively compare our methods with other state-of-the-art methods, including traditional optimization-based method (RLD), supervised learning-based method (VCD-Net) and existing NeRF-based method (DINER). Unlike RLD and DINER, which can perform reconstruction using only light field images, VCD-Net requires the ground-truth as a supervisory signal for training and has poor generalization capability. Therefore, we only compared it with our method on synthetic dataset. Additionally, we conduct extensive experiments to evaluate the superiority of our methods compared with another EFR-based method (3DGS). Note that we introduced wave optics modeling in the rendering process of 3DGS to achieve microscopic light field reconstruction.

\begin{table}[H]
    \centering
    \caption{Comprehensive quantitative comparison according to PSNR, SSIM, LPIPS and running time on dynamic data (immune cells, 1000 frames). The best results are in red and the second best results are in blue.}
    \begin{tabularx}{8.8cm}{XXXXX}
        \toprule
         & VCD-Net & RLD & DINER & Ours \\
        \midrule
        PSNR & 30.04 & 32.87 & \textcolor{blue}{34.02} & \textcolor{red}{38.03} \\
        SSIM & 0.776 & 0.828 & \textcolor{blue}{0.905} & \textcolor{red}{0.935} \\
        LPIPS & 11.08 & 9.919 & \textcolor{blue}{3.801} & \textcolor{red}{2.211} \\
        Time & \textcolor{red}{0.019} & 7.1 & 10.5 & \textcolor{blue}{3.4} \\
        \bottomrule
    \end{tabularx}
    \label{tab_dynamic}
\end{table}

\noindent\textbf{Quantitative comparison on synthetic dataset.} \cref{tab_synthetic} reports the average metrics of all methods on synthetic data. \cref{fig_syn_compare} shows the XY and XZ projection of several slices in 3D intensity volume.

We conducted a comparative analysis of FFT loss against commonly used loss functions in the field of image reconstruction, specifically Perception loss\cite{lpips} and SSIM Loss\cite{ssimloss}, within the mouse neuron scene from the synthetic dataset (at the top of Fig.\ref{fig_syn_compare}). Metrics were calculated for each slice, and the results are presented in Fig.\ref{fig_loss_plot}. It is noteworthy that the presence of defocus aberration significantly complicates the reconstruction of slices that are located farther from the focal plane.
As shown in Fig.\ref{fig_3dgs_compare}, we conducted a qualitative comparison between 3DGS and our method using the USAF resolution test chart. Furthermore, we incorporated FFT Loss into the optimization process of 3DGS and compared all reconstruction results.
Tab.\ref{tab_dynamic} provides a comprehensive quantitative comparison of all methods applied to dynamic data (in the middle of Fig.\ref{fig_syn_compare}), including performance metrics and the average time required to reconstruct each frame following the first frame. To ensure fairness, all the methods are tested on the same PC equipped with an Intel 3.0GHz CPU and an NVIDIA RTX 3090 GPU.
By analyzing the reconstruction results of all methods on the synthetic dataset, we can draw the following conclusions:
\begin{itemize}
    \item[$\bullet$]On the synthetic dataset, our method demonstrates superior reconstruction accuracy, particularly when compared to traditional RLD algorithm. Our method also demonstrates enhanced lateral and axial resolution, particularly in regions with abundant high-frequency details and high cellular density.
    \item[$\bullet$]In comparison to other loss functions utilized for reconstruction, FFT Loss more effectively mitigates the loss of high-frequency information caused by defocus aberrations and the high-frequency cutoff effects introduced by the objective lens. Furthermore, this loss function can be integrated into other approaches, such as 3DGS, in a cost-effective and efficient manner, resulting in a significant enhancement of reconstruction outcomes.
    \item[$\bullet$]Compared to optimization-based methods, our approach demonstrates superior reconstruction efficiency. By employing explicit representation and FFT loss, we can rapidly capture the differences in high-frequency components between consecutive frames and reconstruct the subsequent frame by explicitly modifying the feature vectors. Note that VCD-Net requires less time for reconstruction, but its performance and generalization capabilities are significantly lacking.
\end{itemize}

\noindent\textbf{Qualitative comparisons on real-world dataset.} 
We utilized the 3D intensity distributions obtained from traditional high-resolution TPM as the ground truth and conducted qualitative comparisons of reconstructions. Fig.\ref{fig_real_compare} illustrates the XY and XZ projections of several slices within three 3D intensity distributions.
The GT volume served as a benchmark, demonstrating clear details and structures in all cases. The RLD method exhibited significant deficiencies, particularly in the reconstruction of both mouse neuron scenes and endoplasmic reticulum cells, where critical high-frequency information was lost, resulting in blurred structures, e.g., the dendritic spine structure of neurons. Nevertheless, the RLD algorithm outperformed DINER in terms of high-frequency detail recovery. This is largely due to RLD's reliance on a Poisson distribution prior, which effectively suppresses noise but simultaneously leads to some loss of detail. This is also the reason why researchers continue to favor the traditional RLD algorithm in practical biological observations.

Our method consistently surpassed both RLD and DINER through several key factors: we employed FFT loss to enhance the recovery of high-frequency information, incorporated Z-TV loss to effectively suppress noise encountered in real biological scenarios, and utilized multi-view information during the reconstruction process to correct aberrations, thereby mitigating degradation in reconstruction quality associated with deep tissue imaging. This combination of techniques underscores the superior capability of our approach in reconstructing intricate biological scenes and highlights its potential applicability in biomedical imaging.

\begin{figure*}[ht]
    \centering
    \includegraphics[width=1\linewidth]{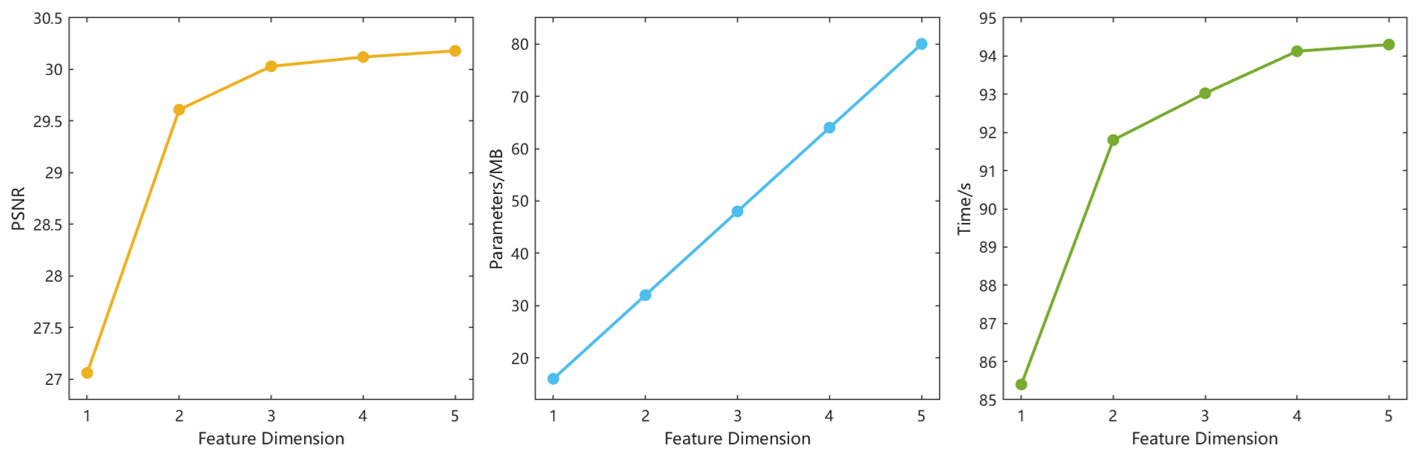}
     \begin{center}
     \vspace{-0.2cm}
  \end{center}  
  \vspace{-0.8cm}
  \caption{The results of an ablation study analyzing the effect of feature dimensions on three critical performance metrics: PSNR, number of parameters, and processing time.}
  \label{fig_feature}
  \vspace{-0.2cm}
\end{figure*}

\noindent\textbf{Digital Adaptive Optics.} 
As shown in Fig.\ref{fig_syn_DAO}, the introduction of artificial aberrations leads to significant blurring and reduced contrast, creating substantial challenges for accurate reconstruction. Nevertheless, our hardware-free aberration correction method effectively mitigates the resolution degradation caused by aberrations compared to other methods, creating a clear progression from the reconstructed outputs to the high-fidelity reference image (TPM). 
The corrected center slice of intensity volume demonstrates a substantial increase in resolution, showcasing enhanced detail and structural clarity compared to the initial low-resolution center view. Furthermore, our approach notably improves the XZ projections by significantly reducing axial motion artifacts and ghosting effects, thus providing clearer representation of structural details and more accurate depth perception.

Fig.\ref{fig_real_DAO} illustrates the effectiveness of DAO on real-world data. We utilize the images captured by TPM under aberration-free conditions as high-fidelity reference images. Observing with 2pSAM under non-water-immersion conditions introduces noticeable aberrations, leading to blurred features and reduced contrast in the center view, particularly in the highlighted area. When our method is applied without DAO, the reconstructed result exhibits marked information loss and blurriness. In contrast, enabling DAO results in significant improvements, as the reconstructed outcomes exhibit sharper details while noticeably alleviating the issue of information loss. This is particularly evident in the highlighted area, where the connections between neurons are clearly visible. 
Overall, the implementation of DAO not only mitigates the detrimental effects of aberrations but also removes the necessity for expensive hardware to achieve effective outcomes in real-world scenarios.

\begin{figure}[h]
    \centering
    \includegraphics[width=1\linewidth]{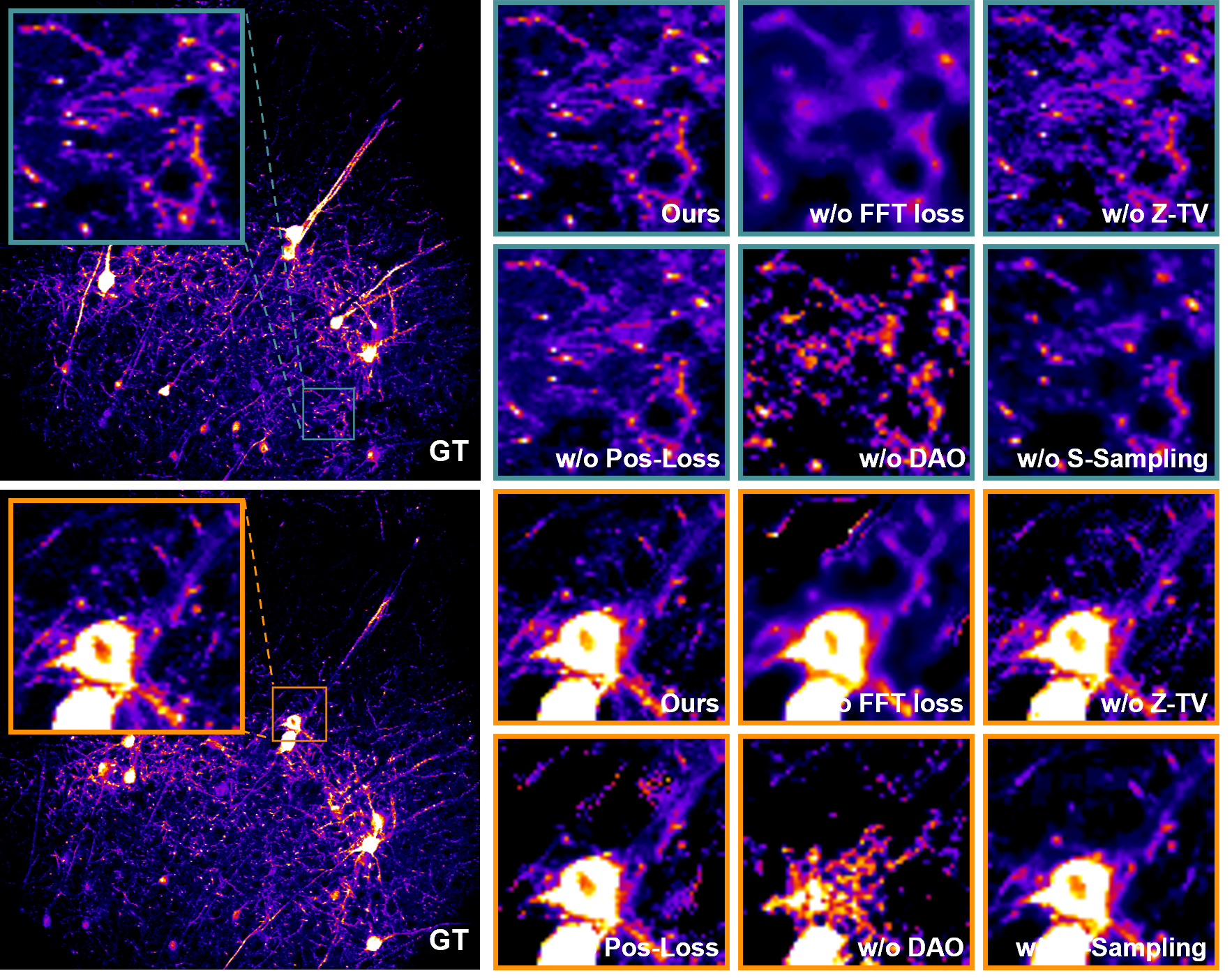}
   \vspace{-0.6cm}
   \caption{The results of an ablation study that evaluates the contributions of all components in our reconstruction method. The left column shows the ground truth, while the right column presents alternative configurations that exclude specific components.}
   \vspace{-0.4cm}
   \label{fig_ablation}
\end{figure}

\begin{table}[H]
    \centering
    \caption{Ablation study on FFT Loss, Z-TV Loss, Positive Loss, DAO and Super-Sampling (SSP) on neurons scene.}
    \begin{tabularx}{8.8cm}{XXXXXXX}
        \toprule
          & Ours & w/o FFT & w/o Z-TV & w/o Pos & w/o DAO & w/o SSP \\
        \midrule
        PSNR & 51.59 & 47.74 & 50.51 & 47.01 & 40.35 & 49.35 \\
        SSIM & 0.983 & 0.964 & 0.974 & 0.946 & 0.915 & 0.963 \\
        LPIPS & 0.468 & 1.533 & 0.881 & 1.141 & 5.753 & 0.901 \\
        \bottomrule
    \end{tabularx}
    \label{tab_ablation}
\end{table}

\subsection{Ablation Study}    
%fig10,11,tab3
As shown in Fig.\ref{fig_feature}, we investigated the impact of the feature dimension $C$ on reconstruction performance, model scomplexity, and efficiency on synthetic data. As $C$ increases, the PSNR shows a significant upward trend, particularly between dimensions 2 and 3, indicating that higher dimensions better capture data characteristics and enhance reconstruction quality. Concurrently, the number of model parameters increases linearly with the feature dimension, which implies that higher dimensions lead to greater model complexity. Additionally, processing time rises from approximately 85 seconds to 94 seconds, suggesting that while the increase is modest, it still impacts computational efficiency. This indicates that when $C>3$, the performance shows only marginal improvements, while model complexity and efficiency still experience a notable increase. Given these considerations, selecting a feature dimension of 3 emerges as a balanced choice. 

Fig.\ref{fig_ablation} reveals the significant contributions of each component to the overall reconstruction performance. The removal of FFT Loss results in a noticeable reduction in detail, leading to blurred images due to the loss of high-frequency information. Similarly, omitting Z-TV Loss increases noise levels and disrupts the structural coherence of the images. When Positive Loss is excluded, negative values may frequently arise during optimization, particularly in areas with low sample intensity, resulting in degraded image quality. The absence of DAO further exacerbates this issue, causing a decline in reconstruction clarity and a marked increase in information loss. Finally, eliminating Super-Sampling leads to a slight loss of detail and overall image sharpness. Collectively, these findings underscore the critical importance of each component in enhancing the reconstruction quality. Tab.\ref{tab_ablation} quantitatively delineates the contributions of all components to the reconstruction results. The results indicate that FFT Loss and DAO exert the most substantial influence on the overall performance.

\section{CONCLUSIONS}
\noindent \textbf{Limitations and Future Work:} 
Our method demonstrates significant potential in brain science, yet it faces several limitations. Although PNR exhibits a notable speed advantage over iterative optimization methods, it still lags behind supervised learning networks, making it challenging to meet the demands for real-time reconstruction. 
Additionally, the method heavily relies on physical modeling of the optical system, discrepancies between the ideal and actual conditions can lead to a decline in reconstruction accuracy. 
Lastly, in scenarios with severe noise, ours requires preprocessing with a denoising network; without this step, reconstruction performance would inevitably degrade.
Future work should focus on several key directions: Firstly, utilizing CUDA toolkit to accelerate the simulated forward projection of LFM could enhance the reconstruction speed. 
Secondly, further improvements in the accuracy of physical modeling are necessary to enhance the precision and reliability of reconstruction results. 
Finally, it is feasible to conduct joint optimization of both the denoising network and the reconstruction network to improve the reconstruction quality in noisy scenarios.

\noindent \textbf{Conclusion:} 
This study introduces the PNR method, which enhances LFM reconstruction quality by combining unsupervised explicit feature representation (EPR) with physical information. Unlike traditional INR, which uses a single MLP, we utilize EPR to represent sample intensity. The explicit feature planes are fully discritized so that each feature vector can be optimized independently in parallel to improve the reconstruction efficiency and eliminate floater artifacts.
Leveraging the influence of aberrations on forward projections from all angles, we can simultaneously perform volume reconstruction and aberration correction during end-to-end optimization. This simplifies the integration of various formulations into the unsupervised LFM framework and demonstrates its potential for incorporating additional task-specific constraints.
Additionally, PNR incorporates the FFT Loss that enhances high-frequency information recovery. By integrating FFT loss, PNR ensures balanced training across frequency bands, leading to improved detail and clarity in reconstructed images. 

Overall, PNR achieves substantial improvements in both reconstruction efficiency and spatial resolution compared to other SOTA methods, particularly in handling high-frequency information and correcting aberrations. 
These innovations position PNR favorably for long-term observation and high-resolution imaging applications. 
% use section* for acknowledgment
\ifCLASSOPTIONcompsoc
  % The Computer Society usually uses the plural form
  \section*{Acknowledgments}
  This work was supported by National Key R\&D Program of China (2023YFB3209700), National Natural Science Foundation of China (62322110, 62071271), and Beijing Natural Science Foundation (JQ21012).
  
\else
  % regular IEEE prefers the singular form
  \section*{Acknowledgment}
\fi

\ifCLASSOPTIONcaptionsoff
  \newpage
\fi

% references section

% can use a bibliography generated by BibTeX as a .bbl file
% BibTeX documentation can be easily obtained at:
% http://mirror.ctan.org/biblio/bibtex/contrib/doc/
% The IEEEtran BibTeX style support page is at:
% http://www.michaelshell.org/tex/ieeetran/bibtex/
\bibliographystyle{IEEEtran}
% argument is your BibTeX string definitions and bibliography database(s)
\bibliography{ref}

% Generated by IEEEtran.bst, version: 1.14 (2015/08/26)
\begin{thebibliography}{10}
\providecommand{\url}[1]{#1}
\csname url@samestyle\endcsname
\providecommand{\newblock}{\relax}
\providecommand{\bibinfo}[2]{#2}
\providecommand{\BIBentrySTDinterwordspacing}{\spaceskip=0pt\relax}
\providecommand{\BIBentryALTinterwordstretchfactor}{4}
\providecommand{\BIBentryALTinterwordspacing}{\spaceskip=\fontdimen2\font plus
\BIBentryALTinterwordstretchfactor\fontdimen3\font minus \fontdimen4\font\relax}
\providecommand{\BIBforeignlanguage}[2]{{%
\expandafter\ifx\csname l@#1\endcsname\relax
\typeout{** WARNING: IEEEtran.bst: No hyphenation pattern has been}%
\typeout{** loaded for the language `#1'. Using the pattern for}%
\typeout{** the default language instead.}%
\else
\language=\csname l@#1\endcsname
\fi
#2}}
\providecommand{\BIBdecl}{\relax}
\BIBdecl

\bibitem{prevedel2014simultaneous}
R.~Prevedel, Y.-G. Yoon, M.~Hoffmann, N.~Pak, G.~Wetzstein, S.~Kato, T.~Schr{\"o}del, R.~Raskar, M.~Zimmer, E.~S. Boyden \emph{et~al.}, ``Simultaneous whole-animal 3d imaging of neuronal activity using light-field microscopy,'' \emph{Nature methods}, vol.~11, no.~7, pp. 727--730, 2014.

\bibitem{lu2024combining}
Z.~Lu, J.~Wu, and Q.~Dai, ``Combining line confocal and scanning light field microscopy to achieve high-resolution observations in deep tissues,'' in \emph{Imaging Systems and Applications}.\hskip 1em plus 0.5em minus 0.4em\relax Optica Publishing Group, 2024, pp. JM4A--11.

\bibitem{pegard2016compressive}
N.~C. P{\'e}gard, H.-Y. Liu, N.~Antipa, M.~Gerlock, H.~Adesnik, and L.~Waller, ``Compressive light-field microscopy for 3d neural activity recording,'' \emph{Optica}, vol.~3, no.~5, pp. 517--524, 2016.

\bibitem{levoy2006light}
M.~Levoy, R.~Ng, A.~Adams, M.~Footer, and M.~Horowitz, ``Light field microscopy,'' in \emph{Acm siggraph 2006 papers}, 2006, pp. 924--934.

\bibitem{wu2021iterative}
J.~Wu, Z.~Lu, D.~Jiang, Y.~Guo, H.~Qiao, Y.~Zhang, T.~Zhu, Y.~Cai, X.~Zhang, K.~Zhanghao \emph{et~al.}, ``Iterative tomography with digital adaptive optics permits hour-long intravital observation of 3d subcellular dynamics at millisecond scale,'' \emph{Cell}, vol. 184, no.~12, pp. 3318--3332, 2021.

\bibitem{LFcao}
X.~Hua, Y.~Wang, S.~Wang, X.~Zou, Y.~Zhou, L.~Li, F.~Yan, X.~Cao, S.~Xiao, D.~P. Tsai \emph{et~al.}, ``Ultra-compact snapshot spectral light-field imaging,'' \emph{Nature communications}, vol.~13, no.~1, p. 2732, 2022.

\bibitem{guo2019fourier}
C.~Guo, W.~Liu, X.~Hua, H.~Li, and S.~Jia, ``Fourier light-field microscopy,'' \emph{Optics express}, vol.~27, no.~18, pp. 25\,573--25\,594, 2019.

\bibitem{rush3d}
Y.~Zhang, M.~Wang, Q.~Zhu, Y.~Guo, B.~Liu, J.~Li, X.~Yao, C.~Kong, Y.~Zhang, Y.~Huang \emph{et~al.}, ``Long-term mesoscale imaging of 3d intercellular dynamics across a mammalian organ,'' \emph{Cell}, 2024.

\bibitem{lu2023virtual}
Z.~Lu, Y.~Liu, M.~Jin, X.~Luo, H.~Yue, Z.~Wang, S.~Zuo, Y.~Zeng, J.~Fan, Y.~Pang \emph{et~al.}, ``Virtual-scanning light-field microscopy for robust snapshot high-resolution volumetric imaging,'' \emph{Nature Methods}, vol.~20, no.~5, pp. 735--746, 2023.

\bibitem{qiao2024zero}
C.~Qiao, Y.~Zeng, Q.~Meng, X.~Chen, H.~Chen, T.~Jiang, R.~Wei, J.~Guo, W.~Fu, H.~Lu \emph{et~al.}, ``Zero-shot learning enables instant denoising and super-resolution in optical fluorescence microscopy,'' \emph{Nature Communications}, vol.~15, no.~1, p. 4180, 2024.

\bibitem{deepsemi}
G.~Zhang, X.~Li, Y.~Zhang, X.~Han, X.~Li, J.~Yu, B.~Liu, J.~Wu, L.~Yu, and Q.~Dai, ``Bio-friendly long-term subcellular dynamic recording by self-supervised image enhancement microscopy,'' \emph{Nature Methods}, vol.~20, no.~12, pp. 1957--1970, 2023.

\bibitem{li2021unsupervised}
X.~Li, G.~Zhang, H.~Qiao, F.~Bao, Y.~Deng, J.~Wu, Y.~He, J.~Yun, X.~Lin, H.~Xie \emph{et~al.}, ``Unsupervised content-preserving transformation for optical microscopy,'' \emph{Light: Science \& Applications}, vol.~10, no.~1, p.~44, 2021.

\bibitem{seg}
M.~Weigert, U.~Schmidt, R.~Haase, K.~Sugawara, and G.~Myers, ``Star-convex polyhedra for 3d object detection and segmentation in microscopy,'' in \emph{Proceedings of the IEEE/CVF winter conference on applications of computer vision}, 2020, pp. 3666--3673.

\bibitem{deepwonder}
Y.~Zhang, G.~Zhang, X.~Han, J.~Wu, Z.~Li, X.~Li, G.~Xiao, H.~Xie, L.~Fang, and Q.~Dai, ``Rapid detection of neurons in widefield calcium imaging datasets after training with synthetic data,'' \emph{Nature Methods}, vol.~20, no.~5, pp. 747--754, 2023.

\bibitem{RLN}
Y.~Li, Y.~Su, M.~Guo, X.~Han, J.~Liu, H.~D. Vishwasrao, X.~Li, R.~Christensen, T.~Sengupta, M.~W. Moyle \emph{et~al.}, ``Incorporating the image formation process into deep learning improves network performance,'' \emph{Nature Methods}, vol.~19, no.~11, pp. 1427--1437, 2022.

\bibitem{vcdnet}
Z.~Wang, L.~Zhu, H.~Zhang, G.~Li, C.~Yi, Y.~Li, Y.~Yang, Y.~Ding, M.~Zhen, S.~Gao \emph{et~al.}, ``Real-time volumetric reconstruction of biological dynamics with light-field microscopy and deep learning,'' \emph{Nature methods}, vol.~18, no.~5, pp. 551--556, 2021.

\bibitem{laasmaa2011application}
M.~Laasmaa, M.~Vendelin, and P.~Peterson, ``Application of regularized richardson-lucy algorithm for deconvolution of confocal microscopy images,'' \emph{Biophysical Journal}, vol. 100, no.~3, p. 139a, 2011.

\bibitem{park2019deepsdf}
J.~J. Park, P.~Florence, J.~Straub, R.~Newcombe, and S.~Lovegrove, ``Deepsdf: Learning continuous signed distance functions for shape representation,'' in \emph{Proceedings of the IEEE/CVF conference on computer vision and pattern recognition}, 2019, pp. 165--174.

\bibitem{mildenhall2021nerf}
B.~Mildenhall, P.~P. Srinivasan, M.~Tancik, J.~T. Barron, R.~Ramamoorthi, and R.~Ng, ``Nerf: Representing scenes as neural radiance fields for view synthesis,'' \emph{Communications of the ACM}, vol.~65, no.~1, pp. 99--106, 2021.

\bibitem{barron2021mip}
J.~T. Barron, B.~Mildenhall, M.~Tancik, P.~Hedman, R.~Martin-Brualla, and P.~P. Srinivasan, ``Mip-nerf: A multiscale representation for anti-aliasing neural radiance fields,'' in \emph{Proceedings of the IEEE/CVF international conference on computer vision}, 2021, pp. 5855--5864.

\bibitem{decaf}
R.~Liu, Y.~Sun, J.~Zhu, L.~Tian, and U.~S. Kamilov, ``Recovery of continuous 3d refractive index maps from discrete intensity-only measurements using neural fields,'' \emph{Nature Machine Intelligence}, vol.~4, no.~9, pp. 781--791, 2022.

\bibitem{DINER}
H.~Zhu, S.~Xie, Z.~Liu, F.~Liu, Q.~Zhang, Y.~Zhou, Y.~Lin, Z.~Ma, and X.~Cao, ``Disorder-invariant implicit neural representation,'' \emph{IEEE Transactions on Pattern Analysis and Machine Intelligence}, 2024.

\bibitem{3dgs}
B.~Kerbl, G.~Kopanas, T.~Leimk{\"u}hler, and G.~Drettakis, ``3d gaussian splatting for real-time radiance field rendering.'' \emph{ACM Trans. Graph.}, vol.~42, no.~4, pp. 139--1, 2023.

\bibitem{2pSAM}
Z.~Zhao, Y.~Zhou, B.~Liu, J.~He, J.~Zhao, Y.~Cai, J.~Fan, X.~Li, Z.~Wang, Z.~Lu \emph{et~al.}, ``Two-photon synthetic aperture microscopy for minimally invasive fast 3d imaging of native subcellular behaviors in deep tissue,'' \emph{Cell}, vol. 186, no.~11, pp. 2475--2491, 2023.

\bibitem{ng2005light}
R.~Ng, M.~Levoy, M.~Br{\'e}dif, G.~Duval, M.~Horowitz, and P.~Hanrahan, ``Light field photography with a hand-held plenoptic camera,'' Ph.D. dissertation, Stanford university, 2005.

\bibitem{refocus}
Y.~Wang, J.~Yang, Y.~Guo, C.~Xiao, and W.~An, ``Selective light field refocusing for camera arrays using bokeh rendering and superresolution,'' \emph{IEEE Signal Processing Letters}, vol.~26, no.~1, pp. 204--208, 2018.

\bibitem{efficientrefocus}
C.~Zhang, G.~Hou, Z.~Zhang, Z.~Sun, and T.~Tan, ``Efficient auto-refocusing for light field camera,'' \emph{Pattern Recognition}, vol.~81, pp. 176--189, 2018.

\bibitem{cheng2021lightSR}
Z.~Cheng, Z.~Xiong, C.~Chen, D.~Liu, and Z.-J. Zha, ``Light field super-resolution with zero-shot learning,'' in \emph{Proceedings of the IEEE/CVF conference on computer vision and pattern recognition}, 2021, pp. 10\,010--10\,019.

\bibitem{wang2022depth}
Y.~Wang, L.~Wang, G.~Wu, J.~Yang, W.~An, J.~Yu, and Y.~Guo, ``Disentangling light fields for super-resolution and disparity estimation,'' \emph{IEEE Transactions on Pattern Analysis and Machine Intelligence}, vol.~45, no.~1, pp. 425--443, 2022.

\bibitem{opal}
P.~Li, J.~Zhao, J.~Wu, C.~Deng, Y.~Han, H.~Wang, and T.~Yu, ``Opal: Occlusion pattern aware loss for unsupervised light field disparity estimation,'' \emph{IEEE Transactions on Pattern Analysis and Machine Intelligence}, 2023.

\bibitem{OAVC}
K.~Han, W.~Xiang, E.~Wang, and T.~Huang, ``A novel occlusion-aware vote cost for light field depth estimation,'' \emph{IEEE transactions on pattern analysis and machine intelligence}, vol.~44, no.~11, pp. 8022--8035, 2021.

\bibitem{jin2020deep}
J.~Jin, J.~Hou, J.~Chen, H.~Zeng, S.~Kwong, and J.~Yu, ``Deep coarse-to-fine dense light field reconstruction with flexible sampling and geometry-aware fusion,'' \emph{IEEE Transactions on Pattern Analysis and Machine Intelligence}, vol.~44, no.~4, pp. 1819--1836, 2020.

\bibitem{lfreview}
C.~Yi, L.~Zhu, D.~Li, and P.~Fei, ``Light field microscopy in biological imaging,'' \emph{Journal of Innovative Optical Health Sciences}, vol.~16, no.~01, p. 2230017, 2023.

\bibitem{broxton2013wave}
M.~Broxton, L.~Grosenick, S.~Yang, N.~Cohen, A.~Andalman, K.~Deisseroth, and M.~Levoy, ``Wave optics theory and 3-d deconvolution for the light field microscope,'' \emph{Optics express}, vol.~21, no.~21, pp. 25\,418--25\,439, 2013.

\bibitem{fish1995blind}
D.~Fish, A.~Brinicombe, E.~Pike, and J.~Walker, ``Blind deconvolution by means of the richardson--lucy algorithm,'' \emph{JOSA A}, vol.~12, no.~1, pp. 58--65, 1995.

\bibitem{lu2019phase}
Z.~Lu, J.~Wu, H.~Qiao, Y.~Zhou, T.~Yan, Z.~Zhou, X.~Zhang, J.~Fan, and Q.~Dai, ``Phase-space deconvolution for light field microscopy,'' \emph{Optics express}, vol.~27, no.~13, pp. 18\,131--18\,145, 2019.

\bibitem{zhao2022sparse}
W.~Zhao, S.~Zhao, L.~Li, X.~Huang, S.~Xing, Y.~Zhang, G.~Qiu, Z.~Han, Y.~Shang, D.-e. Sun \emph{et~al.}, ``Sparse deconvolution improves the resolution of live-cell super-resolution fluorescence microscopy,'' \emph{Nature biotechnology}, vol.~40, no.~4, pp. 606--617, 2022.

\bibitem{zipfel2003nonlinear}
W.~R. Zipfel, R.~M. Williams, and W.~W. Webb, ``Nonlinear magic: multiphoton microscopy in the biosciences,'' \emph{Nature biotechnology}, vol.~21, no.~11, pp. 1369--1377, 2003.

\bibitem{cahalan2002two}
M.~D. Cahalan, I.~Parker, S.~H. Wei, and M.~J. Miller, ``Two-photon tissue imaging: seeing the immune system in a fresh light,'' \emph{Nature Reviews Immunology}, vol.~2, no.~11, pp. 872--880, 2002.

\bibitem{kong2023neuron}
C.~Kong, Y.~Wang, and G.~Xiao, ``Neuron populations across layer 2-6 in the mouse visual cortex exhibit different coding abilities in the awake mice,'' \emph{Frontiers in Cellular Neuroscience}, vol.~17, p. 1238777, 2023.

\bibitem{TPM}
F.~Helmchen and W.~Denk, ``Deep tissue two-photon microscopy,'' \emph{Nature methods}, vol.~2, no.~12, pp. 932--940, 2005.

\bibitem{TPMprinciples}
K.~Svoboda and R.~Yasuda, ``Principles of two-photon excitation microscopy and its applications to neuroscience,'' \emph{Neuron}, vol.~50, no.~6, pp. 823--839, 2006.

\bibitem{icha2017phototoxicity}
J.~Icha, M.~Weber, J.~C. Waters, and C.~Norden, ``Phototoxicity in live fluorescence microscopy, and how to avoid it,'' \emph{BioEssays}, vol.~39, no.~8, p. 1700003, 2017.

\bibitem{patterson2000photobleaching}
G.~H. Patterson and D.~W. Piston, ``Photobleaching in two-photon excitation microscopy,'' \emph{Biophysical journal}, vol.~78, no.~4, pp. 2159--2162, 2000.

\bibitem{hampson2021adaptive}
K.~M. Hampson, R.~Turcotte, D.~T. Miller, K.~Kurokawa, J.~R. Males, N.~Ji, and M.~J. Booth, ``Adaptive optics for high-resolution imaging,'' \emph{Nature Reviews Methods Primers}, vol.~1, no.~1, p.~68, 2021.

\bibitem{wu2022aberration}
J.~Wu, Y.~Guo, C.~Deng, A.~Zhang, H.~Qiao, Z.~Lu, J.~Xie, L.~Fang, and Q.~Dai, ``An integrated imaging sensor for aberration-corrected 3d photography,'' \emph{Nature}, vol. 612, no. 7938, pp. 62--71, 2022.

\bibitem{cao2024aberration}
Z.~Cao, N.~Li, L.~Zhu, J.~Wu, Q.~Dai, and H.~Qiao, ``Aberration-robust monocular passive depth sensing using a meta-imaging camera,'' \emph{Light: Science \& Applications}, vol.~13, no.~1, p. 236, 2024.

\bibitem{ji2010adaptive}
N.~Ji, D.~E. Milkie, and E.~Betzig, ``Adaptive optics via pupil segmentation for high-resolution imaging in biological tissues,'' \emph{Nature methods}, vol.~7, no.~2, pp. 141--147, 2010.

\bibitem{verinaz2021deep}
H.~Verinaz-Jadan, P.~Song, C.~L. Howe, P.~Quicke, A.~J. Foust, and P.~L. Dragotti, ``Deep learning for light field microscopy using physics-based models,'' in \emph{2021 IEEE 18th International Symposium on Biomedical Imaging (ISBI)}.\hskip 1em plus 0.5em minus 0.4em\relax IEEE, 2021, pp. 1091--1094.

\bibitem{ying2023parf}
H.~Ying, B.~Jiang, J.~Zhang, D.~Xu, T.~Yu, Q.~Dai, and L.~Fang, ``Parf: Primitive-aware radiance fusion for indoor scene novel view synthesis,'' in \emph{Proceedings of the IEEE/CVF International Conference on Computer Vision}, 2023, pp. 17\,706--17\,716.

\bibitem{wang2021neus}
P.~Wang, L.~Liu, Y.~Liu, C.~Theobalt, T.~Komura, and W.~Wang, ``Neus: Learning neural implicit surfaces by volume rendering for multi-view reconstruction,'' \emph{arXiv preprint arXiv:2106.10689}, 2021.

\bibitem{liu2024finer}
Z.~Liu, H.~Zhu, Q.~Zhang, J.~Fu, W.~Deng, Z.~Ma, Y.~Guo, and X.~Cao, ``Finer: Flexible spectral-bias tuning in implicit neural representation by variable-periodic activation functions,'' in \emph{Proceedings of the IEEE/CVF Conference on Computer Vision and Pattern Recognition}, 2024, pp. 2713--2722.

\bibitem{pamir}
Z.~Zheng, T.~Yu, Y.~Liu, and Q.~Dai, ``Pamir: Parametric model-conditioned implicit representation for image-based human reconstruction,'' \emph{IEEE transactions on pattern analysis and machine intelligence}, vol.~44, no.~6, pp. 3170--3184, 2021.

\bibitem{qiao2022neuphysics}
Y.-L. Qiao, A.~Gao, and M.~Lin, ``Neuphysics: Editable neural geometry and physics from monocular videos,'' \emph{Advances in Neural Information Processing Systems}, vol.~35, pp. 12\,841--12\,854, 2022.

\bibitem{wanghx}
H.~Wang, T.~Yu, T.~Yang, H.~Qiao, and Q.~Dai, ``Neural physical simulation with multi-resolution hash grid encoding,'' in \emph{Proceedings of the AAAI Conference on Artificial Intelligence}, vol.~38, no.~6, 2024, pp. 5410--5418.

\bibitem{he2024mmpi}
Y.~He, P.~Wang, Y.~Hu, W.~Zhao, R.~Yi, Y.-J. Liu, and W.~Wang, ``Mmpi: a flexible radiance field representation by multiple multi-plane images blending,'' in \emph{2024 IEEE International Conference on Robotics and Automation (ICRA)}.\hskip 1em plus 0.5em minus 0.4em\relax IEEE, 2024, pp. 15\,395--15\,401.

\bibitem{costvolume}
A.~Yu, W.~Guo, B.~Liu, X.~Chen, X.~Wang, X.~Cao, and B.~Jiang, ``Attention aware cost volume pyramid based multi-view stereo network for 3d reconstruction,'' \emph{ISPRS Journal of Photogrammetry and Remote Sensing}, vol. 175, pp. 448--460, 2021.

\bibitem{muller2022instant}
T.~M{\"u}ller, A.~Evans, C.~Schied, and A.~Keller, ``Instant neural graphics primitives with a multiresolution hash encoding,'' \emph{ACM transactions on graphics (TOG)}, vol.~41, no.~4, pp. 1--15, 2022.

\bibitem{chen2022tensorf}
A.~Chen, Z.~Xu, A.~Geiger, J.~Yu, and H.~Su, ``Tensorf: Tensorial radiance fields,'' in \emph{European conference on computer vision}.\hskip 1em plus 0.5em minus 0.4em\relax Springer, 2022, pp. 333--350.

\bibitem{zernike}
K.~Rahbar, K.~Faez, and E.~Attaran~Kakhki, ``Phase wavefront aberration modeling using zernike and pseudo-zernike polynomials,'' \emph{Journal of the Optical Society of America A}, vol.~30, no.~10, pp. 1988--1993, 2013.

\bibitem{lpips}
R.~Zhang, P.~Isola, A.~A. Efros, E.~Shechtman, and O.~Wang, ``The unreasonable effectiveness of deep features as a perceptual metric,'' in \emph{Proceedings of the IEEE conference on computer vision and pattern recognition}, 2018, pp. 586--595.

\bibitem{leackyrelu}
B.~Xu, ``Empirical evaluation of rectified activations in convolutional network,'' \emph{arXiv preprint arXiv:1505.00853}, 2015.

\bibitem{ssp}
C.~Wang, X.~Wu, Y.-C. Guo, S.-H. Zhang, Y.-W. Tai, and S.-M. Hu, ``Nerf-sr: High quality neural radiance fields using supersampling,'' in \emph{Proceedings of the 30th ACM International Conference on Multimedia}, 2022, pp. 6445--6454.

\bibitem{zhangyi}
Y.~Zhang, Y.~Wang, M.~Wang, Y.~Guo, X.~Li, Y.~Chen, Z.~Lu, J.~Wu, X.~Ji, and Q.~Dai, ``Multi-focus light-field microscopy for high-speed large-volume imaging,'' \emph{PhotoniX}, vol.~3, no.~1, p.~30, 2022.

\bibitem{deepcad}
X.~Li, G.~Zhang, J.~Wu, Y.~Zhang, Z.~Zhao, X.~Lin, H.~Qiao, H.~Xie, H.~Wang, L.~Fang \emph{et~al.}, ``Reinforcing neuron extraction and spike inference in calcium imaging using deep self-supervised denoising,'' \emph{Nature methods}, vol.~18, no.~11, pp. 1395--1400, 2021.

\bibitem{deepcadrt}
X.~Li, Y.~Li, Y.~Zhou, J.~Wu, Z.~Zhao, J.~Fan, F.~Deng, Z.~Wu, G.~Xiao, J.~He \emph{et~al.}, ``Real-time denoising enables high-sensitivity fluorescence time-lapse imaging beyond the shot-noise limit,'' \emph{Nature Biotechnology}, vol.~41, no.~2, pp. 282--292, 2023.

\bibitem{pnevmatikakis2017normcorre}
E.~A. Pnevmatikakis and A.~Giovannucci, ``Normcorre: An online algorithm for piecewise rigid motion correction of calcium imaging data,'' \emph{Journal of neuroscience methods}, vol. 291, pp. 83--94, 2017.

\bibitem{ssimloss}
H.~Zhao, O.~Gallo, I.~Frosio, and J.~Kautz, ``Loss functions for image restoration with neural networks,'' \emph{IEEE Transactions on computational imaging}, vol.~3, no.~1, pp. 47--57, 2016.

\end{thebibliography}
%
% <OR> manually copy in the resultant .bbl file
% set second argument of \begin to the number of references
% (used to reserve space for the reference number labels box)
% \begin{thebibliography}{1}

% \bibitem{IEEEhowto:kopka}
% H.~Kopka and P.~W. Daly, \emph{A Guide to \LaTeX}, 3rd~ed.\hskip 1em plus
%   0.5em minus 0.4em\relax Harlow, England: Addison-Wesley, 1999.

% \end{thebibliography}

% biography section
% 
% If you have an EPS/PDF photo (graphicx package needed) extra braces are
% needed around the contents of the optional argument to biography to prevent
% the LaTeX parser from getting confused when it sees the complicated
% \includegraphics command within an optional argument. (You could create
% your own custom macro containing the \includegraphics command to make things
% simpler here.)

\vfill
\begin{IEEEbiography}[{\includegraphics[width=1in,height=1.25in,clip,keepaspectratio]{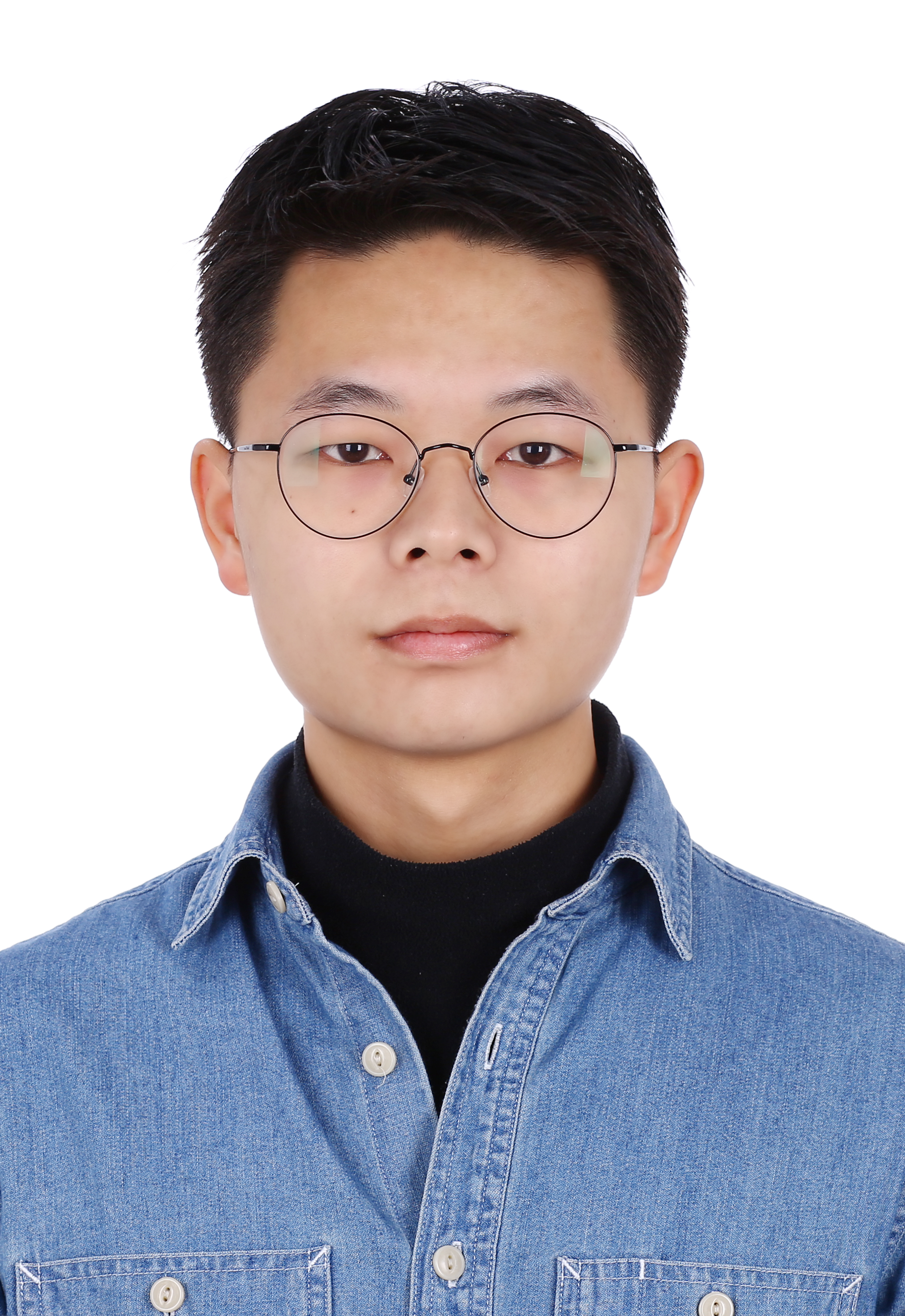}}]{Jiayin Zhao} is a graduate student in Department of Automation, Tsinghua University. He received the B.S. degree in Measurement and Control from Beihang University, China, in 2020. His research interest includes microscopy imaging and 3D reconstruction.
\end{IEEEbiography}
\begin{IEEEbiography}[{\includegraphics[width=1in,height=1.25in,clip,keepaspectratio]{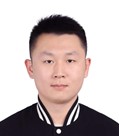}}]{Zhifeng Zhao} is currently a postdoctoral researcher in the Department of Automation at Tsinghua University. He earned his Bachelor's degree in Mechanical Design, Manufacturing, and Automation from Central South University, China, in 2017, and completed his Ph.D. in Automation at Tsinghua University in 2024. His research interests focus on multiphoton microscopy and computational imaging.
\end{IEEEbiography}
\begin{IEEEbiography}[{\includegraphics[width=1in,height=1.25in,clip,keepaspectratio]{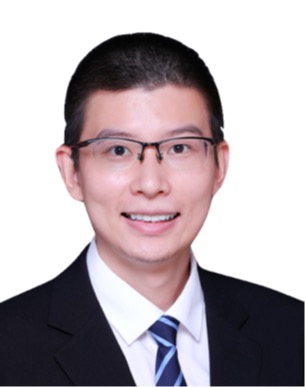}}]{Jiamin Wu} received the B.S.and Ph.D.degrees in automation from Tsinghua University, Beijing, China, in 2014 and 2019, respectively. He is currently an associate professor with the Department of Automation, Tsinghua University. His research interests include computational microscopy and optical computing, with a particular emphasis on developing computation-based optical setups for observing large-scale biological dynamics in vivo. Dr. Wu was an Associate Editor for IEEE TRANSACTIONS ON CIRCUITS AND SYSTEMS FOR VIDEO TECHNOLOGY, and a Reviewer of Light: Science \& Applications, Optica, and Optics Express.
\end{IEEEbiography}
\begin{IEEEbiography}[{\includegraphics[width=1in,height=1.25in,clip,keepaspectratio]{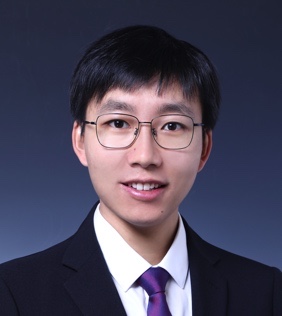}}]{Tao Yu}  received a B.S. degree in Measurement and Control from Hefei University of Technology, in 2012, and a Ph.D. degree from Beihang University, in 2019. From 2020 to 2022, he worked as a postdoc with the Department of Automation, Tsinghua University, China. He is currently an associate researcher of BNRist, Tsinghua University. His research area lies at the intersection between computer vision, deep learning, and computer graphics. His main research focus is digitizing humans using different kinds of sensors and algorithms.
\end{IEEEbiography}
\begin{IEEEbiography}[{\includegraphics[width=1in,height=1.25in,clip,keepaspectratio]{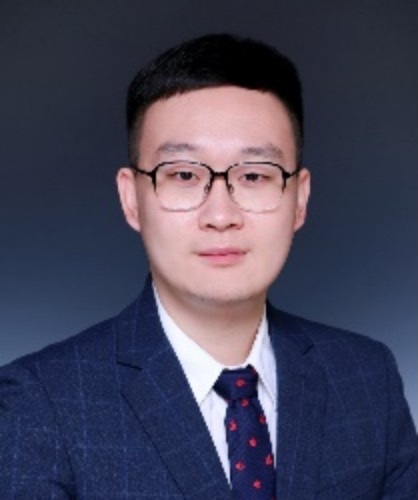}}]{Hui Qiao} received the BE degree in automation from Tsinghua University, China, in 2013, and the Ph.D degree in control science and engineering from Tsinghua University, China, in 2019. He is currently an associate professor at Tsinghua University. His research interests include computational imaging, computational microscopy, and machine learning.
\end{IEEEbiography}
\vfill

% or if you just want to reserve a space for a photo:

% You can push biographies down or up by placing
% a \vfill before or after them. The appropriate
% use of \vfill depends on what kind of text is
% on the last page and whether or not the columns
% are being equalized.

%\vfill

% Can be used to pull up biographies so that the bottom of the last one
% is flush with the other column.
%\enlargethispage{-5in}

% that's all folks
\end{document}